\documentclass[a4paper,11pt]{article}
\pdfoutput=1 

\usepackage{jheppub} 

\usepackage{times}
\usepackage[T1]{fontenc} 
\usepackage[utf8]{inputenc}
\usepackage{dsfont}
\usepackage{arydshln}
\usepackage{slashed}
\usepackage{xcolor}
\usepackage{multirow}
\usepackage{graphicx}
\usepackage{cleveref}
\usepackage{hyperref}
\usepackage{mathdots}
\usepackage{physics}
\usepackage{subfigure}
\usepackage{tikz}
\usepackage{soul}
\usetikzlibrary{decorations.pathreplacing}

\def\D{\Delta}

\def\e{\epsilon}

\def\m{\mu}
\def\n{\nu}
\def\s{\sigma}
\def\r{\rho}

\def\avg#1{\langle #1 \rangle}

\makeatletter
\newsavebox\myboxA
\newsavebox\myboxB
\newlength\mylenA
\newcommand*\xoverline[2][0.75]{%
    \sbox{\myboxA}{$\m@th#2$}%
    \setbox\myboxB\null
    \ht\myboxB=\ht\myboxA%
    \dp\myboxB=\dp\myboxA%
    \wd\myboxB=#1\wd\myboxA
    \sbox\myboxB{$\m@th\overline{\copy\myboxB}$}
    \setlength\mylenA{\the\wd\myboxA}
    \addtolength\mylenA{-\the\wd\myboxB}%
    \ifdim\wd\myboxB<\wd\myboxA%
       \rlap{\hskip 0.5\mylenA\usebox\myboxB}{\usebox\myboxA}%
    \else
        \hskip -0.5\mylenA\rlap{\usebox\myboxA}{\hskip 0.5\mylenA\usebox\myboxB}%
    \fi}
\makeatother

\crefname{table}{Table}{Tables}
\crefname{equation}{Eq.}{Eqs.}
\crefname{appendix}{App.}{Apps.}
\crefname{section}{Sec.}{Secs.}
\crefname{figure}{Fig.}{Figs.}

\newcommand{\be}{\begin{eqnarray}}
\newcommand{\ee}{\end{eqnarray}}

\def\ie{\emph{i.e.}\,\,}
\def\etc{\emph{etc.}\,\,}
\def\eg{\emph{e.g.}\,\,}

\def\Lag{\mathcal{L}}

\title{2, 12, 117, 1959, 45171, 1170086, ...: A Hilbert series for the QCD chiral Lagrangian}

\author[a]{Luk\'a\v{s} Gr\'af,}\emailAdd{lukas.graf@mpi-hd.mpg.de}
\author[b]{Brian Henning,}\emailAdd{brian.henning@unige.ch}
\author[c]{Xiaochuan Lu,}\emailAdd{xlu@uoregon.edu}
\author[d]{Tom Melia,}\emailAdd{tom.melia@ipmu.jp}
\author[d,e,f,1]{Hitoshi Murayama\note{Hamamatsu Professor}}\emailAdd{hitoshi@berkeley.edu}\emailAdd{hitoshi.murayama@ipmu.jp}

\affiliation[a]{Max-Planck-Institut f\"{u}r Kernphysik, Saupfercheckweg 1, 69117 Heidelberg, Germany}
\affiliation[b]{D\'epartment de Physique Th\'eorique, Universit\'e de Gen\`{e}ve, 24 quai Ernest-Ansermet, 1211 Gen\`eve 4, Switzerland}
\affiliation[c]{Institute for Fundamental Science, Department of Physics, University of Oregon, Eugene, OR 97403, USA}
\affiliation[d]{Kavli Institute for the Physics and Mathematics of the
  Universe (WPI), University of Tokyo Institutes for Advanced Study, University of Tokyo,
  Kashiwa 277-8583, Japan}
\affiliation[e]{Department of Physics, University of California, Berkeley, CA 94720, USA}
\affiliation[f]{Ernest Orlando Lawrence Berkeley National Laboratory, Berkeley, CA 94720, USA}

\abstract{We apply Hilbert series techniques to the enumeration of operators in the mesonic QCD chiral Lagrangian. Existing Hilbert series technologies for non-linear realizations are extended to incorporate the external fields. The action of charge conjugation is addressed by folding the $\frak{su}(n)$ Dynkin diagrams, which we detail in an appendix that can be read separately as it has potential broader applications. New results include the enumeration of anomalous operators appearing in the chiral Lagrangian at order $p^8$, as well as enumeration of $CP$-even, $CP$-odd, $C$-odd, and $P$-odd terms beginning from order $p^6$. The method is extendable to very high orders, and we present results up to order $p^{16}$.
\newline
\newline

{\small
\noindent (The title sequence is the number of independent $C$-even \emph{and} $P$-even operators in the mesonic QCD chiral Lagrangian with three light flavors of quarks, at chiral dimensions $p^2$, $p^4$, $p^6$, ...)
}
}

\begin{document}
\maketitle
\flushbottom
\newpage

\section{Introduction}
\label{sec:intro}

The appearance of Hilbert series in the particle physics literature began with their application to counting gauge invariants in supersymmetric theories~\cite{Benvenuti:2006qr,Feng:2007ur,Gray:2008yu}, and flavour invariants~\cite{Jenkins:2009dy,Hanany:2010vu}, and they were subsequently  established for the purpose of enumerating Lorentz invariant operators that can appear in the Lagrangian of an effective field theory (EFT)~\cite{Lehman:2015via, Henning:2015daa, Lehman:2015coa, Henning:2015alf, Henning:2017fpj,Ruhdorfer:2019qmk} (see~\cite{Kobach:2017xkw} for non-relativistic EFTs). One application of particular significance is to the Standard Model (SM) EFT~\cite{Weinberg:1979sa, Weinberg:1980bf, Buchmuller:1985jz}---Hilbert series systematize the enumeration of SMEFT operators~\cite{Henning:2015alf, Marinissen:2020jmb}. The SMEFT has as its constituents massless fields that transform linearly under the gauge symmetries. These two properties  enable a rigorous treatment of the operator redundancies coming from Equations of Motion (EOM) and Integration by Parts (IBP) identities via conformal representation theory, as shown in~\cite{Henning:2017fpj}.

In this paper, we demonstrate that Hilbert series can similarly systematize the enumeration of operators in the mesonic QCD chiral Lagrangian~\cite{Pagels:1974se, Weinberg:1978kz, Gasser:1983yg, Gasser:1984gg}. The endeavour to enumerate/construct operators in this EFT parallels that in the SMEFT. Much effort has gone into constructing operator bases at higher order in the EFT expansion---the chiral dimension $p^k$ in this case. Since the leading order $p^2$ and next-to-leading order $p^4$ terms in the chiral Lagrangian were computed in the original works~\cite{Gasser:1983yg, Gasser:1984gg}, results at order $p^6$ have appeared~\cite{Fearing:1994ga, Bijnens:1999sh, Bijnens:2001bb, Ebertshauser:2001nj, Haefeli:2007ty, Colangelo:2012ipa}, and recently at order $p^8$~\cite{Bijnens:2018lez}. Parallels also exist whereby operator redundancies (due to EOM, IBP, or symmetry group relations) were missed in some of the earlier attempts at order $p^6$ (see~\cite{Bijnens:2018lez} for a review of the details), providing a compelling reason to also have a systematic approach.

The Hilbert series technology for EFT operator enumeration has been expounded in some detail in the literature (we refer the interested reader to \eg \cite{Lehman:2015via,Henning:2017fpj}). However, for an application to the chiral Lagrangian, it is necessary to make some generalizations and technical advances. First, a Hilbert series approach for non-linearly realized global symmetries was developed in \cite{Henning:2017fpj}. This was rooted in the CCWZ formalism \cite{Coleman:1969sm, Callan:1969sn}, and only pion operators were considered. On the other hand, the QCD chiral Lagrangian community uses a slightly modified setup where external source fields are introduced \cite{Gasser:1983yg,Gasser:1984gg}, allowing one to extend the global $SU(N_f)_L \times SU(N_f)_R$ symmetry into a \textit{local} symmetry. This introduces additional building blocks beyond those discussed in \cite{Henning:2017fpj}, which must be incorporated; see \cref{sec:ChiralLag}. Similar to the pion field discussed in \cite{Henning:2017fpj}, some of these external fields also do not form conformal representations (reps), precluding a rigorous and straightforward treatment of IBP redundancies via conformal representation theory. We follow \cite{Henning:2017fpj} and use ideas from the theory of differential forms to systematically address IBP relations.

The second technical advance we need is to systematically incorporate the charge conjugation $C$ into the enumeration of the operator basis. The bulk of the chiral Lagrangian community focuses on \emph{both} $C$-even \emph{and} $P$-even operators \cite{Fearing:1994ga, Bijnens:1999sh, Bijnens:2001bb, Ebertshauser:2001nj, Haefeli:2007ty, Colangelo:2012ipa, Bijnens:2018lez}. The reason for this is phenomenological: $CP$ violation in the QCD Lagrangian is \textit{small}, appearing in the phase of the quark mass matrix and the $\theta$ term. Of course, these lead to important physical phenomena like $K$-$\bar{K}$ mixing and $K_L\rightarrow \pi^0 \nu \bar{\nu}$ decay. However, it is generally assumed that the smallness of these terms in the UV Lagrangian (the QCD Lagrangian) justifies keeping only the leading terms in the IR Lagrangian (the chiral Lagrangian), so that one can safely ignore higher-dimension operators that violate $C$ and/or $P$. It was shown in~\cite{Henning:2017fpj} how Hilbert series can capture the effect of parity $P$ transformations, \eg so as to separately enumerate $P$-even and $P$-odd operators in a Lagrangian. There is a beautiful mirroring of the treatment of the action of $P$ developed in~\cite{Henning:2017fpj} in how $C$ is treated in the current work. While parity acts as an outer automorphism of the Lie algebra of the Euclidean spacetime symmetry group $SO(d)$, $C$ acts as an outer automorphism on the Lie algebra of the unbroken $SU(N_f)_V$ symmetry group of the chiral Lagrangian. The construction of a Hilbert series in both cases follows from the notion of `folding' a Dynkin diagram, explored in detail in Appendix C of~\cite{Henning:2017fpj} for the action of $P$, and in \cref{appsec:folding} of the current paper for $C$.

In this paper, we extend the existing Hilbert series technology, and apply it to the mesonic chiral Lagrangian. We reproduce/confirm all up-to-date operator enumeration results that we are aware of in the literature. We also extend them to higher orders and obtain new results. Among the $C$-even \emph{and} $P$-even operators, the chiral Lagrangian community often distinguishes operators which lead to processes where the \textit{intrinsic} parity of the process changes, while $P$ is still nevertheless conserved, such as the process $\pi \pi \to \pi \pi \pi$ which involves an \textit{odd} numbers of pions. In practice, such operators in the Lagrangian will have an $\epsilon$-tensor so that the total operator remains $P$-even. These operators are termed ``anomalous'' by the chiral Lagrangian community \cite{Ebertshauser:2001nj, Bijnens:2001bb, Colangelo:2012ipa}. In light of this, our most immediately relevant new results in this paper are the enumeration of anomalous operators at chiral dimension $p^8$, which supplements the non-anomalous sector results in \cite{Bijnens:2018lez}, and hence completes the list of \emph{both} $C$-even \emph{and} $P$-even operators.

In addition, our method  provides enumeration of other sectors of operators, such as the $CP$-even, $CP$-odd, $C$-odd, and $P$-odd ones. We are not aware of previous results in the literature starting at dimension $p^6$, and we provide the operator content of these sectors in this work. In the Standard Model, $CP$ violation is so particularly small that it is a great laboratory for new physics effects. In fact, even mass dimension eight SMEFT operators that are suppressed by multi-TeV scales can be important. From the chiral Lagrangian point of view, they are encoded by operators of higher chiral dimensions that include flavor-violating spurions ($\Sigma$ in this paper). If there are light particles from new physics, even higher dimension operators may play a role.

We emphasize that our ``full'' results are the Hilbert series themselves, containing maximum information about the operator content which is much more useful for the actual \emph{construction} of operators. Different sectors of operators are just various components or combinations of them (see \cref{sec:results} for details). For this purpose, we include the Hilbert series at order $p^6$ and $p^8$ as an auxiliary material that accompanies this paper, and encourage the interested reader to investigate the accompanying {\tt Mathematica} notebook. We also emphasize that our method is completely systematic, which we illustrate by applying it to count operators up to order $p^{16}$.

As well as being used to describe the low-energy limit of QCD, chiral Lagrangians are used in many models of physics beyond the Standard Model. Perhaps the first examples are the technicolor models \cite{Weinberg:1975gm,Susskind:1978ms} where the electroweak symmetry breaking is described by the chiral Lagrangian. In this case, the Nambu-Goldstone Bosons (NGB) are eaten by the $W$ and $Z$ bosons without a Higgs boson. Even though such models are widely believed to be ruled out experimentally, in particular by the measurements of the oblique electroweak parameters $S$ and $T$ \cite{Peskin:1990zt,Peskin:1991sw}, it would be an interesting question to ask whether higher order operators in the chiral Lagrangian would ameliorate the tension with precision electroweak data. In this case, the observed Higgs boson would appear as an extra non-NGB degree of freedom. Its description would require the so-called Higgs Effective Field Theory (HEFT) \cite{Feruglio:1992wf} which we would like to discuss elsewhere \cite{HilbertHEFT}. On the other hand, if the observed Higgs boson is regarded to be one of the Nambu-Goldstone bosons, the model is a composite Higgs model \cite{Giudice:2007fh}. Such models are well-motivated as they can explain the hierarchy problem by protecting the Higgs boson mass against large quadratic divergences. One of the main difficulties, however, is to obtain a large enough Higgs mass because the self-coupling vanishes for Nambu-Goldstone bosons if the symmetry is exact; again higher order operators can be of interest on this question. Finally, there are also applications of chiral Lagrangians to study dark matter candidates, such as Strongly-Interacting Massive Particles (SIMPs) \cite{Hochberg:2014kqa}, where the dark matter freezes out in a $3\rightarrow 2$ annihilation process via the Wess-Zumino term in the chiral Lagrangian. The mass spectrum among dark matter particles can be sensitive to higher order operators \cite{Tsai:2020vpi}. In all, classifying operators in chiral Lagrangians can be an important problem.

The structure of the paper is as follows. \cref{sec:ChiralLag} serves to outline the notation and terminology we use throughout the paper,  and reviews the form of the linearly transforming fields that were introduced in~\cite{Bijnens:1999sh,Bijnens:2001bb} for use in the construction of the Lagrangian. In Section~\ref{sec:method} we provide the details of how a Hilbert series based on these building blocks is constructed, with particular emphasis on how this is constructed on the different $C$ and $P$ odd and even branches.  Finally, \cref{sec:results} presents information contained within the Hilbert series in various ways, for example coarse-grained enumeration of operators, breakdown by their $C$ and $P$ transformations \etc

We include four appendices, and one auxiliary file. \cref{appsec:SPM} provides explicit character formulae that enter the Hilbert series for the various fields in the chiral Lagrangian on the different $C$ and $P$ branches. \cref{appsec:folding} contains information on how the character formulae on the $C$ odd branch are obtained from `folding'  Dynkin diagrams of the special unitary group. \cref{appsec:p8} gives a more detailed breakdown of the new results that enumerate the operators appearing in the anomalous Lagrangian at chiral dimension $p^8$. \cref{appsec:nf45} provides enumeration of operators for four and five flavours of light quark up to chiral dimension 16. The auxiliary file {\tt Hilbert-series-p6-and-p8.nb} provides the full Hilbert series for the chiral Lagrangian at chiral dimension $p^6$ and $p^8$.

\section{Linear building blocks of the chiral Lagrangian}
\label{sec:ChiralLag}

In this section, we briefly review the setup of the chiral Lagrangian. Following \cite{Gasser:1983yg,Gasser:1984gg} (see e.g. \cite{Bijnens:2018lez} for the notation we use in the following), we consider the UV theory as the QCD Lagrangian with four external source fields---vector $v_\mu$, axial-vector $a_\mu$, scalar $s$, and pseudo-scalar $p$:
\begin{equation}
\Lag_\text{UV} = \Lag_\text{QCD} + \bar{q}\gamma^\mu\left(v_\mu+a_\mu\gamma^5\right) q - \bar{q} \left(s-ip\gamma^5\right) q \,.
\end{equation}
The quark field $q$ has $N_f$ components (flavors). The external fields are real $N_f\times N_f$ matrices due to hermiticity of the Lagrangian. In addition, $v_\mu$ and $a_\mu$ are assumed to be traceless. With these external fields, the global chiral symmetry $\left(g_L, g_R\right) \in SU(N_f)_L\times SU(N_f)_R$ satisfied by QCD can be extended into a local one. The consequently required transformation properties of the external fields are most recognizable in terms of the following combinations
\begin{subequations}
\begin{alignat}{2}
\ell_\mu  &\equiv v_\mu - a_\mu  &&\quad\longrightarrow\quad  g_L\, \ell_\mu\, g_L^\dagger - i\left(\partial_\mu g_L\right)g_L^\dagger \,, \\[5pt]
r_\mu  &\equiv v_\mu + a_\mu  &&\quad\longrightarrow\quad  g_R\, r_\mu\, g_R^\dagger - i\left(\partial_\mu g_R\right)g_R^\dagger \,, \\[5pt]
\Sigma &\equiv -\frac{2}{N_f F_\pi^2} \langle\bar{q}q\rangle (s+ip)  &&\quad\longrightarrow\quad  g_R\, \Sigma\, g_L^\dagger \,,
\end{alignat}
\end{subequations}
with $F_\pi$ denoting the pion decay constant.

In the IR, the chiral symmetry is spontaneously broken to $SU(N_f)_V$ by the quark bilinear vev $\langle\bar{q}q\rangle$. The basic building block of the resulting EFT is the Goldstone matrix field $\xi(\pi)$, which transforms nonlinearly as
\begin{equation}
\xi \to g_R\, \xi\, h^{-1} \left(\xi, g_L, g_R\right) = h\left(\xi, g_L, g_R\right)\, \xi\, g_L^{-1} \,,
\end{equation}
with a certain element in the unbroken group $h\left(\xi, g_L, g_R\right) \in SU(N_f)_V$ that also depends on the field $\xi$. Employing the linearization recipe proposed by CCWZ \cite{Coleman:1969sm,Callan:1969sn}, one can find the linearly transforming building blocks under the unbroken group $SU(N_f)_V$ (see \eg \cite{Bijnens:1999sh,Bijnens:2001bb,Bijnens:2018lez}):
\begin{subequations}\label{eqn:LinearBlocks}
\begin{alignat}{2}
u_\mu &= u_\mu^a T^a &&\equiv i \left[\xi^\dagger\left(\partial_\mu-ir_\mu\right)\xi - \xi\left(\partial_\mu-i\ell_\mu\right)\xi^\dagger\right] \,,\\[5pt]
\Sigma_\pm + \langle\Sigma_\pm\rangle &= \Sigma_\pm^a T^a + \langle\Sigma_\pm\rangle \mathbf{1} &&\equiv \xi^\dagger\, \Sigma\, \xi^\dagger \pm \xi\, \Sigma^\dagger\, \xi \,, \\[5pt]
f_\pm^{\mu\nu} &= f_\pm^{\mu\nu,a} T^a &&\equiv \xi\, F_L^{\mu\nu}\, \xi^\dagger \pm \xi^\dagger\, F_R^{\mu\nu}\, \xi \,,
\end{alignat}
\end{subequations}
with $T^a$ denoting the $SU(N_f)_V$ generators in the fundamental representation. Here $F_{L/R}^{\mu\nu}$ are the field strengths for $\ell^\mu$/$r^\mu$:
\begin{subequations}
\begin{align}
F_L^{\mu\nu} &= \partial^\mu \ell^\nu -\partial^\nu \ell^\mu -i\comm{\ell^\mu}{\ell^\nu} \,,\\[5pt]
F_R^{\mu\nu} &= \partial^\mu r^\nu -\partial^\nu r^\mu -i\comm{r^\mu}{r^\nu} \,.
\end{align}
\end{subequations}
In the second line of \cref{eqn:LinearBlocks}, we have split the field into the trace part $\langle\Sigma_\pm\rangle$ and the traceless part $\Sigma_\pm$ for future convenience.

\begin{table}[t]
\renewcommand{\arraystretch}{2.0}
\setlength{\arrayrulewidth}{.3mm}
\setlength{\tabcolsep}{1 em}
\begin{center}
\begin{tabular}{c|cccc}
Fields                     & $SU(N_f)_V$ & Intrinsic Parity & Charge Conjugation         & Chiral Dim\\\hline
$u_\mu$                    & adjoint     & $-$              & $u_\mu^T$                  & 1 \\
$\Sigma_\pm$               & adjoint     & $\pm$            & $\Sigma_\pm^T$             & 2 \\
$\langle\Sigma_\pm\rangle$ & singlet     & $\pm$            & $\langle\Sigma_\pm\rangle$ & 2 \\
$f_{\pm\mu\nu}$            & adjoint     & $\pm$            & $\mp f_{\pm\mu\nu}^T$      & 2
\end{tabular}\vspace{0.2cm}
\caption{Transformation properties and chiral dimensions of linear building blocks of the chiral Lagrangian (see \eg \cite{Bijnens:2018lez}).}
\label{tbl:Transformations}
\end{center}
\end{table}

To build the chiral Lagrangian, we are interested in the effective operators built by the fields in \cref{eqn:LinearBlocks} together with the covariant derivative $D_\mu$, which are invariant under the Lorentz $SO(4)$ symmetry,\footnote{Throughout this paper, we work in Euclidean spacetime where the Lorentz symmetry is $SO(4)$.} internal unbroken $SU(N_f)_V$ symmetry, parity $P$, as well as charge conjugation $C$.\footnote{As usual, if one is interested in finding an operator basis, there are of course linear redundancies to remove, such as EOM and IBP.} One also needs a power counting scheme to truncate the EFT expansion---the so-called \emph{chiral dimension} in the case of the chiral Lagrangian. For the linear building blocks in \cref{eqn:LinearBlocks}, the chiral dimensions are respectively $\left\{u_\mu, \Sigma_\pm, \langle\Sigma_\pm\rangle, f_{\pm\mu\nu}\right\} \longrightarrow \left\{1, 2, 2, 2\right\}$. In addition, each power of covariant derivative has chiral dimension one. We summarize the transformation properties and chiral dimensions of the linear building blocks $\phi=\left\{u_\mu, \Sigma_\pm, \langle\Sigma_\pm\rangle, f_{\pm\mu\nu}\right\}$ in \cref{tbl:Transformations} (see \eg \cite{Bijnens:2018lez}).

\section{Hilbert series for the chiral Lagrangian}
\label{sec:method}

In this section, we briefly summarize the procedure of using Hilbert series to find the operator basis of the chiral Lagrangian. The Hilbert series method is a systematic approach that is explored in some detail in \cite{Henning:2017fpj}. In this section, we will keep the general discussion brief and focus on its special features when applied to the case of the chiral Lagrangian.

We compute the main part of the Hilbert series $H_0$ as
\begin{equation}
H_0 (\phi, p) = \int\dd\mu_\text{Internal}^{}(y) \int\dd\mu_\text{Spacetime}^{}(x)\, \frac{1}{P(p, x)}\, Z(\phi, p, x, y) \,.
\label{eqn:H0}
\end{equation}
The components of this expression are briefly explained in order:
\begin{enumerate}
  \item The set of all the local operators modulo the EOM redundancies forms a linear space, which furnishes a representation of spacetime and internal symmetry groups. The integrand $Z(\phi, p, x, y)$  (closely related to a partition function) is what is known as a {\it character} (a trace over a group matrix) of this (highly reducible) representation, further graded by $\phi$ and $p$. It can be computed as
      \begin{equation}
      Z(\phi, p, x, y) = \prod_i\frac{1}{\det\left[1 - \phi_i\, g_i(p, x, y)\right]} = \exp \left[ \sum_i \sum_{n=1}^\infty \frac{1}{n} \phi_i^n \tr \left(g_i^n\right) \right] \,.
      \label{eqn:Zexpression}
      \end{equation}
      Here $\phi=\left\{u, \Sigma_\pm, \langle\Sigma_\pm\rangle, f_{\pm}\right\}$ collectively denotes spurion variables that represent all the linear building blocks (fields) of the chiral Lagrangian; $p$ is the power counting parameter, whose power indicates the chiral dimension of the term; $x$ and $y$ are variables for the character function (\ie trace) of the operator's representation matrix under the spacetime and internal symmetries, respectively.

      The representation matrix $g_i(p, x, y)$ of a {\it single particle module}~\cite{Henning:2017fpj} (defined as $\phi_i$ and its derivatives) is a tensor product of that for the spacetime symmetry group and that for the internal symmetry group:
      \begin{equation}
      g_i(p, x, y) = g_i^\text{Spacetime}(p, x) \otimes g_i^\text{Internal}(y) \,.
      \label{eqn:gtensor}
      \end{equation}
      For the case of the chiral Lagrangian, the spacetime symmetry group is the Lorentz $SO(4)$, and a ${\mathbb Z}_2$ group $\mathcal{P}=\{1, P\}$ due to parity; the internal symmetry group is the unbroken $SU(N_f)_V$, and a ${\mathbb Z}_2$ group $\mathcal{C}=\{1, C\}$ due to charge conjugation. The two \(\mathbb{Z}_2\) actions do not commute with their respective groups, so the underlying group structure is the semi-direct product groups \(SO(4) \rtimes \mathcal{P}\) and \(SU(N_f)_V \rtimes \mathcal{C}\), see Sec.~\ref{subsec:P_and_C}. The character variable $x$ parameterizes a maximal torus of the spacetime symmetry group $SO(4)$, $x=\left(x_1, x_2\right)$ with two being the rank of $SO(4)$.
      The $y$ variable has a similar structure. Eigenvalues of the representation matrix $g$ are integer powers of the character variables. When these eigenvalues all come with the trivial overall sign (\ie plus), we have
      \begin{equation}
      \chi(z) \equiv \tr\Big(g(z)\Big)  \quad\Longrightarrow\quad  \tr \left(g^n\right) = \chi\left(z^n\right) \,.
      \label{eqn:trpowers}
      \end{equation}
      Here we use $z$ to denote a generic character variable, and have adopted an abbreviated notation $z^n\equiv\left(z_1^n, \cdots, z_r^n\right)$. Making use of this and the factorization in \cref{eqn:gtensor}, we get
      \begin{equation}
      \tr \Big( g_i^n(p, x, y) \Big) = \chi_i^\text{Spacetime}(p^n, x^n)\, \chi_i^\text{Internal}(y^n) \,.
      \label{eqn:TracePowers}
      \end{equation}
      Therefore, we can better organize \cref{eqn:Zexpression} into
      \begin{equation}
      Z(\phi, p, x, y) = \exp \left[ \sum_i \sum_{n=1}^\infty \frac{1}{n}\, \chi_i\left(\phi_i^n, p^n, x^n, y^n\right) \right] \,,
      \label{eqn:Zcharacter}
      \end{equation}
      with $\chi_i$ the graded character for each single particle module:
      \begin{equation}
      \chi_i\left(\phi_i, p, x, y\right) \equiv \phi_i\, \chi_i^\text{Spacetime}(p, x)\, \chi_i^\text{Internal}(y) \,.
      \end{equation}
      In \cref{appsec:SPM}, we discuss the single particle module formed by each field $\phi_i$ (and its covariant derivatives), and provide the character list $\chi_i^\text{Spacetime}(p, x)$ in \cref{eqn:chilistSpacetime} and $\chi_i^\text{Internal}(y)$ in \cref{eqn:chilistInternal}.
  \item The integral $\int\dd\mu_\text{Spacetime}^{}(x)\, \frac{1}{P(p,x)}$ takes care of imposing the spacetime symmetries, including Lorentz $SO(4)$ invariance, translation invariance (namely IBP redundancies), as well as parity (if desired). When parity is not imposed, this integral is simply
      \begin{equation}
      \int\dd\mu_\text{Spacetime}^{}(x)\, \frac{1}{P(p, x)} = \int\dd\mu_{SO(4)}^{}(x)\, \frac{1}{P_+(p, x)} \,,
      \label{eqn:Poincare}
      \end{equation}
      with
      \begin{equation}
      P_+(p, x) = \frac{1}{\left(1 - p x_1\right) \left(1 - p x_1^{-1}\right) \left(1 - p x_2\right) \left(1 - p x_2^{-1}\right)} \,.
      \label{eqn:Pplus}
      \end{equation}
      Because of the orthonormality of characters, the Haar measure integral $\int\dd\mu_{SO(4)}^{}(x)$ ({\it i.e.} integral over the group $SO(4)$) selects out the Lorentz representations of our interest. For example, without the factor $\frac{1}{P(p,x)}$, this would select out the Lorentz singlets (scalars) out of the operator space represented by $Z$, and hence `imposes' the Lorentz symmetry. The role of the additional factor $\frac{1}{P(p,x)}$ is to remove the IBP redundancies (equivalently, imposing translation invariance). See \cref{subsec:IBP} below for more explanations.
  \item The Haar measure integral  $\int\dd\mu_\text{Internal}^{}(y)$ takes care of imposing the internal symmetries, including the $SU(N_f)_V$ invariance, as well as the charge conjugation invariance (if desired). When charge conjugation is not imposed, this integral is simply
      \begin{equation}
      \int\dd\mu_\text{Internal}^{}(y) = \int\dd\mu_{SU(N_f)}^{}(y) \,,
      \end{equation}
      which selects out the $SU(N_f)_V$ singlets via character orthonormality.
\end{enumerate}

Clearly, in practical evaluation of the Hilbert series given in \cref{eqn:H0}, we will need the character expressions for various reps, as well as the Haar measures (called Weyl integration formula) for the classical Lie groups. These can be found in many group theory textbooks, \eg \cite{Brocker:2003,FultonHarris}. See also Apps. A and B in \cite{Henning:2017fpj} for summaries.

\subsection{Addressing IBP redundancies}
\label{subsec:IBP}

Without the factor $\frac{1}{P(p,x)}$, the Haar measure integral in \cref{eqn:Poincare} selects out all the scalar ($SO(4)$ singlet) operators. The additional factor $\frac{1}{P(p,x)}$ makes the Hilbert series into an alternating sum of rank-$k$ antisymmetric $SO(4)$ tensors (which we will call \emph{forms} as in \cite{Henning:2015alf,Henning:2017fpj}), starting from $k=0$, namely scalar. This largely removes the IBP redundancies, except for the small caveat due to the existence of co-closed but not co-exact forms \cite{Henning:2015alf,Henning:2017fpj}. In most generality, these forms give a further correction term $\Delta H$ in addition to the main piece $H_0$ in \cref{eqn:H0}, making the total Hilbert series $H=H_0+\Delta H$. (See Section 7 in \cite{Henning:2017fpj} for detailed elaborations.) However, experience has shown that $\Delta H$ only contains operators at relatively low EFT orders. For example, it is proven in \cite{Henning:2017fpj} that $\Delta H$ in SMEFT only contains operators with mass dimension $\text{dim}\le 4$, which follows from conformal representation theory. For an EFT of pions, strong evidence was given in \cite{Henning:2017fpj} that $\Delta H$ only contains operators with mass dimension $\text{dim}\le 4$, and it was conjectured that no operators with $\text{dim}> 4$ contribute to $\Delta H$. For our chiral Lagrangian at hand, we enumerated all the co-closed but not co-exact forms by hand for chiral dimension below or equal to $p^4$, and found that none of them would survive once $C$ and $P$ are both imposed (see \cref{appsec:p4} for a detailed elaboration). Therefore, for \emph{both} $C$-even \emph{and} $P$-even operators, we have $H=H_0$ at $p^2$ and $p^4$. Beyond $p^4$, we \emph{conjecture} that $\Delta H$ does not contribute to the mesonic chiral Lagrangian, even when $C$ and/or $P$ are not imposed. This conjecture is supported by the agreement we found between $H_0$ predictions and the enumerations by other methods in the literature, as well as other supporting evidence given in \cite{Henning:2017fpj}. With this conjecture in mind, we will drop the subscript in $H_0$ from now on, and simply call the expression given in \cref{eqn:H0} $H$.

\subsection{Parity and charge conjugation}
\label{subsec:P_and_C}

\begin{table}[t]
\renewcommand{\arraystretch}{1.8}
\setlength{\arrayrulewidth}{.3mm}
\setlength{\tabcolsep}{1em}
\begin{center}
\begin{tabular}{c|cc}
$O(4)$               & $O_+(4)$                               & $O_-(4)$                                      \\[3pt]
\hline
Haar measure         & $\dd\mu_{SO(4)}(x)$                    & $\dd\mu_{Sp(2)}\left(\tilde{x}\right)$        \\[3pt]
\hline
$\left(l_1,0\right)$ rep character & $\chi_{\left(l_1,0\right)}^{SO(4)}(x)$ & $\eta_P^{}\chi_{l_1}^{Sp(2)}\left(\tilde{x}\right)$ \\[3pt]
$\left(l_1,l_2\ne0\right)$ rep character & $\chi_{\left(l_1,l_2\right)}^{SO(4)}(x) + \chi_{\left(l_1,-l_2\right)}^{SO(4)}(x)$ & $0$ \\
\end{tabular}\vspace{0.2cm}
\caption{Haar measure and characters for $O(4)$ in terms of those of the classical Lie groups. A general unitary $SO(4)$ irreducible rep (irrep) is labelled by its highest weight $l=\left(l_1, l_2\right)$, which satisfies $l\in \frac12 \mathbb{Z}$, $l_1-l_2\in\mathbb{Z}$, and $l_1\ge\left|l_2\right|$. When $l_2=0$, the $SO(4)$ irrep forms an $O(4)$ irrep itself. In this case, there is an overall intrinsic sign choice $\eta_P^{}=\pm$ for the odd branch character, which distinguishes real scalar (or vector \etc) from pseudo-scalar (or pseudo-vector \etc). When $l_2\ne0$, the $SO(4)$ reps $\left(l_1,l_2\right)\oplus\left(l_1,-l_2\right)$ form an $O(4)$ irrep. In this case, the odd branch character vanishes. Our notation $Sp(2k)$ denotes the compact symplectic group, $Sp(2k)\equiv Sp(2k,\mathbb{C})\cap U(2k)$.}
\label{tbl:chiBranchesO4}
\end{center}
\end{table}

A detailed derivation and explanation on how to impose parity via the Hilbert series can be found in App. C of \cite{Henning:2017fpj}. Here we summarize the practical recipe. We promote the Lorentz symmetry $SO(4)$ to the disconnected group by parity $P$ (its outer automorphism): $O(4) = SO(4) \rtimes \mathcal{P} = \left\{O_+(4), O_-(4)\right\}$. Then the $P\text{-even}$ Hilbert series is given by an average over the two disconnected branches:
\begin{align}
H^{P\text{-even}}(\phi, p) &= \int\dd\mu_\text{Internal}^{}(y)\, \frac12 \Bigg[ \int\dd\mu_{O_+(4)}^{}(x)\, \frac{1}{P_+(p, x)}\, Z^{P^+}\left(\phi, p, x, y\right) \notag\\
&\hspace{90pt} + \int\dd\mu_{O_-(4)}^{}\left(\tilde{x}\right)\, \frac{1}{P_-\left(p, \tilde{x}\right)}\, Z^{P^-}\left(\phi, p, \tilde{x}, y\right) \Bigg] \,,
\label{eqn:HPevenDemo}
\end{align}
where the function $P_-\left(p, \tilde{x}\right)$ is
\begin{equation}
P_-\left(p, \tilde{x}\right) = \frac{1}{\left(1 - p x_1\right) \left(1 - p x_1^{-1}\right) \left(1- p^2\right)} \,,
\label{eqn:Pminus}
\end{equation}
and where we introduced the variable $\tilde{x}$ for the odd branch elements $g_-$ to distinguish it from $x$ used for the even branch elements $g_+$, because they have different numbers of components (see \cref{tbl:xVariables}).
To compute the above two branches of Hilbert series, we need the characters of various reps, as well as the Haar measure for the disconnected group $O(4)$, on both its branches $O_\pm(4)$. In \cref{tbl:chiBranchesO4}, we provide a summary of these in terms of those of the classical Lie groups. They can be derived using the folding technique explained in App. C of \cite{Henning:2017fpj}.

Imposing charge conjugation can be achieved in a similar way as imposing parity. In particular, we extend the internal symmetry $SU(N_f)$ to the disconnected orbit group $\widetilde{SU}(N_f)\equiv SU(N_f)\rtimes\mathcal{C} = \left\{\widetilde{SU}_+(N_f), \widetilde{SU}_-(N_f)\right\}$, and the $C\text{-even}$ Hilbert series is given by an average over the two disconnected branches:
\begin{align}
H_{N_f}^{C\text{-even}}(\phi, p) &= \frac12 \Bigg[ \int\dd\mu_{\widetilde{SU}_+(N_f)}^{}(y)\, \int\dd\mu_\text{Spacetime}^{}(x)\, Z_{N_f}^{C^+}\left(\phi, p, x, y\right) \notag\\
&\hspace{15pt} + \int\dd\mu_{\widetilde{SU}_-(N_f)}^{}\left(\tilde{y}\right)\, \int\dd\mu_\text{Spacetime}^{}(x)\, Z_{N_f}^{C^-}\left(\phi, p, x, \tilde{y}\right) \Bigg] \,,
\label{eqn:HCevenDemo}
\end{align}
where again we are using $\tilde{y}$ for the odd branch to distinguish it from $y$ used for the even branch, as they have different numbers of components (see \cref{tbl:xVariables}). To compute these two branches of the Hilbert series, we need the characters of the singlet and adjoint rep, as well as the Haar measure for the disconnected group $\widetilde{SU}(N_f)$, on both of its branches $\widetilde{SU}_\pm(N_f)$. These are summarized in \cref{tbl:chiBranchesGN}, in terms of those of the classical Lie groups. These results can be derived by folding the Dynkin diagram $A_r=\frak{su}(r+1)$ with $r=N_f-1$, which we will explain in \cref{appsec:folding}. Note that in \cref{tbl:chiBranchesGN} we need to distinguish the even $N_f=2k$ and the odd $N_f=2k+1$ cases. In addition, the $SU(N_f)$ adjoint representation is self-conjugate under charge conjugation. In this case, there is an overall intrinsic sign choice $\eta_C^{}=\pm$ for the odd branch character. For the chiral Lagrangian fields listed in \cref{tbl:Transformations}, fields transforming as plus transpose (i.e. $u_\mu$, $\Sigma_\pm$, and $f_{-\mu\nu}$) and those transforming as minus transpose (i.e. $f_{+\mu\nu}$) should obviously take opposite signs $\eta_C^{}$; indeed the first set ($u_\mu$, $\Sigma_\pm$, and $f_{-\mu\nu}$) takes $\eta_C^{}=-1$ and the latter set, i.e. $f_{+\mu\nu}$ takes $\eta_C^{}=+1$.

\begin{table}[t]
\renewcommand{\arraystretch}{1.8}
\setlength{\arrayrulewidth}{.3mm}
\setlength{\tabcolsep}{1em}
\begin{center}
\begin{tabular}{c|ccc}
$\widetilde{SU}(N_f)\equiv SU(N_f)\rtimes\mathcal{C}$ & $\widetilde{SU}_+(N_f)$          & $\widetilde{SU}_-(N_f=2k)$     & $\widetilde{SU}_-(N_f=2k+1)$       \\[3pt]
\hline
Haar measure                             & $\dd\mu_{SU(N)}(y)$              & $\dd\mu_{SO(2k+1)}(\tilde{y})$ & $\dd\mu_{Sp(2k)}(\tilde{y})$\\[3pt]
\hline
singlet rep characters                   & $1$                              & $1$                            & $1$ \\
adjoint rep characters                   & $\chi_\text{adjoint}^{SU(N)}(y)$ & $\eta_C^{}\chi_\text{fundamental}^{SO(2k+1)}\left(\tilde{y}\right)$ & $\eta_C^{}\chi_\text{fundamental}^{Sp(2k)}\left(\tilde{y}\right)$ \\
\end{tabular}\vspace{0.2cm}
\caption{Haar measure and characters for $\widetilde{SU}(N_f)\equiv SU(N_f)\rtimes\mathcal{C}$ in terms of those of the classical Lie groups. $SU(N_f)$ adjoint representation is self-conjugate under charge conjugation. In this case, there is an overall intrinsic sign choice $\eta_C^{}=\pm$ for the odd branch character. Our notation $Sp(2k)$ denotes the compact symplectic group.}
\label{tbl:chiBranchesGN}
\end{center}
\end{table}

\subsection{Character branches}
\label{subsec:CharacterBranches}

It is clear from the discussion above that we need the integrand $Z(\phi, p, x, y)$ on different branches of the disconnected groups: $Z_{N_f}^{C^\pm P^\pm}, Z_{N_f}^{C^\pm P^\mp}$. These are the ($\phi, p$)-graded characters that can be evaluated as in \cref{eqn:Zexpression}, taking the (representation matrix of the) group element $g_{\pm\pm}, g_{\pm\mp}$ according to the branch selected. However, a subtlety is that the expression given in \cref{eqn:Zcharacter} only applies to the fully even branch $Z_{N_f}^{C^+ P^+}$. When an odd branch is involved, \cref{eqn:trpowers} breaks down for even powers $n=2k$, because certain eigenvalues of $g$, which are still integer powers of the character variables, come with a minus sign. Taking the parity case as an example, in the vector rep of $O(4)$ (\ie $\left(l_1, l_2\right) = (1,0)$), the odd element $g_-$ can be diagonalized into
\begin{equation}
g_-\left(\tilde x\right) \quad\longrightarrow\quad \mqty(x_1 & 0 & 0 & 0\\ 0 & x_1^{-1} & 0 & 0\\ 0 & 0 & 1 & 0\\ 0 & 0 & 0 & -1) \,.
\label{eqn:EigenExample}
\end{equation}
The odd branch character is therefore
\begin{equation}
\chi_-\left(\tilde{x}\right) \equiv \tr\Big(g_-\left(\tilde{x}\right)\Big) = x_1 + x_1^{-1} \,.
\end{equation}
This is as expected from the results in \cref{tbl:chiBranchesO4}. However, due to the minus sign in front of the last eigenvalue in \cref{eqn:EigenExample}, we see that the trace of even powers of $g_-$ is less straightforward:
\begin{subequations}\label{eqn:trgminus}
\begin{align}
\tr \Big(g_-^{2k+1}\left(\tilde{x}\right)\Big) &= \chi_-\left(\tilde{x}^{2k+1}\right) \,, \\[10pt]
\tr \Big(g_-^{2k}\left(\tilde{x}\right)\Big) &\ne \chi_-\left(\tilde{x}^{2k}\right) \,.
\end{align}
\end{subequations}
The remedy is actually to use $\chi_+$ instead for even powers:
\begin{equation}
\tr\Big(g_-^{2k}\left(\tilde{x}\right)\Big) = \chi_+\left(\bar{x}^{2k}\right) \,,
\label{eqn:tr2kchiplus}
\end{equation}
with a new variable $\bar{x}$ in place of $\tilde{x}$. This new variable $\bar{x}$ has as many components as the variable $x$, among which the number of independent ones however is only as many as that of $\tilde{x}$. One obtains $\bar{x}$ from $x$ by relating components in accordance with folding the Dynkin diagram. In \cref{tbl:xVariables}, we summarize the relations among $x$, $\tilde{x}$, and $\bar{x}$ for $O(4)$, and $y$, $\tilde{y}$, and $\bar{y}$ for $\widetilde{SU}(N_f)\equiv SU(N_f)\rtimes\mathcal{C}$. This subtlety of evaluating $\tr\left(g^n\right)$ reflected by \cref{eqn:trgminus,eqn:tr2kchiplus} is also summarized in \cref{tbl:TracePowers}.

\begin{table}[t]
\renewcommand{\arraystretch}{2.0}
\setlength{\arrayrulewidth}{.3mm}
\setlength{\tabcolsep}{1 em}
\begin{center}
\begin{tabular}{c|ccc}
            & $O(4)$                  & $\widetilde{SU}(2k)\equiv SU(2k)\rtimes\mathcal{C}$ & $\widetilde{SU}(2k+1)\equiv SU(2k+1)\rtimes\mathcal{C}$ \\
\hline
$x$ or $y$         & $\left(x_1, x_2\right)$ & $\left(y_1, \cdots, y_{2k}\right)$      & $\left(y_1, \cdots, y_{2k+1}\right)$        \\
$\tilde{x}$ or $\tilde{y}$ & $x_1$                   & $\left(y_1, \cdots, y_k\right)$         & $\left(y_1, \cdots, y_k\right)$             \\
$\bar{x}$ or $\bar{y}$   & $\left(x_1,   1\right)$ & $\left(\sqrt{y_1}, \cdots, \sqrt{y_k}, \frac{1}{\sqrt{y_k}}, \cdots, \frac{1}{\sqrt{y_1}}\right)$ & $\left(\sqrt{y_1}, \cdots, \sqrt{y_k}, 1, \frac{1}{\sqrt{y_k}}, \cdots, \frac{1}{\sqrt{y_1}}\right)$
\end{tabular}\vspace{0.2cm}
\caption{Relations among $x$, $\tilde{x}$, and $\bar{x}$ for $O(4)$, and $y$, $\tilde{y}$, and $\bar{y}$ for $\widetilde{SU}(N_f)\equiv SU(N_f)\rtimes\mathcal{C}$. Note that our $y$ variables for $\widetilde{SU}(N_f)$ appear to have one more component than the rank of the group $r=N_f-1$. This is because it is more convenient to use an $(r+1)$-dimensional vector space for the root and weight system of $\frak{su}(r+1)$ where all roots are orthogonal to the vector $(1,1,\cdots,1)$. Consequently, a relation among the $r+1$ components of $y$ is understood: $\prod_{i=1}^{r+1} y_i=1$. The variable $\bar{y}$ is obtained from $y$ by relating components in accordance with folding the Dynkin diagram; see \cref{appsec:folding} for details.}
\label{tbl:xVariables}
\end{center}
\end{table}

\begin{table}[t]
\renewcommand{\arraystretch}{2.0}
\setlength{\arrayrulewidth}{.3mm}
\setlength{\tabcolsep}{1em}
\begin{center}
\begin{tabular}{c|cc}
$G$                   & $G_+$                    & $G_-$  \\\hline
$\tr\left(g^n\right)$ & $\chi_+\left(z^n\right)$ & $\chi_-\left(\tilde{z}^{n=2k+1}\right) \,\,,\,\, \chi_+\left(\bar{z}^{n=2k}\right)$
\end{tabular}\vspace{0.2cm}
\caption{Traces of odd and even powers of group elements $g\in G=\left\{G_+, G_-\right\}$ on the even branch $G_+$ and the odd branch $G_-$. Here we have adopted an abbreviated notation $z^n\equiv\left(z_1^n, \cdots, z_r^n\right)$, and similarly for $\tilde{z}^n$ and $\bar{z}^n$. The group $G$ here could be either the group $O(4)$ or the charge conjugation orbit group $\widetilde{SU}(N_f)\equiv SU(N_f)\rtimes\mathcal{C}$, and the variable $z$ could be either $x$ or $y$ correspondingly.}
\label{tbl:TracePowers}
\end{center}
\end{table}

Due to the subtlety explained above, we split the $Z$ expression in \cref{eqn:Zcharacter} into odd and even powers:
\begin{align}
Z_{N_f}^{C, P\text{ branch}} &= \exp \Bigg[ \sum_i \sum_{k=0}^\infty \frac{1}{2k+1}\, \chi_{i,\,N_f,\text{ odd-power}}^{C, P\text{ branch}} \left(\phi_i^{2k+1}, p^{2k+1}, x^{2k+1}, y^{2k+1}\right) \notag\\[5pt]
&\hspace{40pt} + \sum_i \sum_{k=1}^\infty \frac{1}{2k}\, \chi_{i,\,N_f,\text{ even-power}}^{C, P\text{ branch}} \left(\phi_i^{2k}, p^{2k}, x^{2k}, y^{2k}\right) \Bigg] \,,
\label{eqn:Zsplit}
\end{align}
where $\chi_{i,\,N_f,\text{ odd-power}}^{C, P\text{ branch}}$ and $\chi_{i,\,N_f,\text{ even-power}}^{C, P\text{ branch}}$ are different functions (except on the branch $C^+ P^+$), as summarized in \cref{tbl:SPMCharacterBranches}. The group element characters $\chi_i^{P^\pm}$ and $\chi_{i,\,N_f}^{C^\pm}$ in \cref{tbl:SPMCharacterBranches} can in turn be obtained from \cref{tbl:chiBranchesO4,tbl:chiBranchesGN}, based on the representations formed by the single particle module. It is a bit nontrivial to compute the characters $\chi_i^{P^\pm}$, as one needs to sum over all the components in a single particle module, which typically all live in different representations in \cref{tbl:chiBranchesO4}. In \cref{appsec:SPM}, we provide explicit expressions of $\chi_i^{P^\pm}$ (\cref{eqn:chilistSpacetime}) and $\chi_{i,\,N_f}^{C^\pm}$ (\cref{eqn:chilistInternal}) for each of the single particle modules in the chiral Lagrangian.

\begin{table}[t]
\renewcommand{\arraystretch}{2.0}
\setlength{\arrayrulewidth}{.3mm}
\setlength{\tabcolsep}{1em}
\begin{center}
\begin{tabular}{c|cc}
$\chi_{i,\,N_f}^{}$ & odd-power                                               & even-power                                               \\\hline
$C^+ P^+$ & \multicolumn{2}{c}{$\phi_i\, \chi_i^{P^+}(p, x)\,\chi_{i,\,N_f}^{C^+}(y)$} \\
$C^+ P^-$ & $\phi_i\, \chi_i^{P^-}\left(p, \tilde{x}\right)\,\chi_{i,\,N_f}^{C^+}(y)$ & $\phi_i\, \chi_i^{P^+}\left(p, \bar{x}\right)\,\chi_{i,\,N_f}^{C^+}(y)$ \\
$C^- P^+$ & $\phi_i\, \chi_i^{P^+}(p, x)\,\chi_{i,\,N_f}^{C^-}\left(\tilde{y}\right)$ & $\phi_i\, \chi_i^{P^+}(p, x)\,\chi_{i,\,N_f}^{C^+}\left(\bar{y}\right)$ \\
$C^- P^-$ & $\phi_i\, \chi_i^{P^-}\left(p, \tilde{x}\right)\,\chi_{i,\,N_f}^{C^-}\left(\tilde{y}\right)$ & $\phi_i\, \chi_i^{P^+}\left(p, \bar{x}\right)\,\chi_{i,\,N_f}^{C^+}\left(\bar{y}\right)$
\end{tabular}\vspace{0.2cm}
\caption{Single particle module characters $\chi_{i,\,N_f,\text{ odd-power}}^{C, P\text{ branch}}$ and $\chi_{i,\,N_f,\text{ even-power}}^{C, P\text{ branch}}$ in terms of group element characters $\chi_i^{P^\pm}$ and $\chi_{i,\,N_f}^{C^\pm}$.}
\label{tbl:SPMCharacterBranches}
\end{center}
\end{table}

\subsection{Hilbert series branches and cases}
\label{subsec:HilbertBranches}

Now that we have defined the integrand $Z$ on each branch of the disconnected groups, it is natural to also define the following \emph{Hilbert series branches}:
\begin{subequations}\label{eqn:HBranches}
\begin{align}
H_{N_f}^{C^+ P^+}(\phi, p) &\equiv \int\dd\mu_{\widetilde{SU}_+(N_f)}\left(y\right)         \int\dd\mu_{O_+(4)}\left(x\right)\,         \frac{1}{P_+\left(p, x\right)}\, Z_{N_f}^{C^+ P^+}\left(\phi, p, x, y\right) \,, \\[8pt]
H_{N_f}^{C^+ P^-}(\phi, p) &\equiv \int\dd\mu_{\widetilde{SU}_+(N_f)}\left(y\right)         \int\dd\mu_{O_-(4)}\left(\tilde{x}\right)\, \frac{1}{P_-\left(p, \tilde{x}\right)}\, Z_{N_f}^{C^+ P^-}\left(\phi, p, \tilde{x}, y\right) \,, \\[8pt]
H_{N_f}^{C^- P^+}(\phi, p) &\equiv \int\dd\mu_{\widetilde{SU}_-(N_f)}\left(\tilde{y}\right) \int\dd\mu_{O_+(4)}\left(x\right)\,         \frac{1}{P_+\left(p, x\right)}\, Z_{N_f}^{C^- P^+}\left(\phi, p, x, \tilde{y}\right) \,, \\[8pt]
H_{N_f}^{C^- P^-}(\phi, p) &\equiv \int\dd\mu_{\widetilde{SU}_-(N_f)}\left(\tilde{y}\right) \int\dd\mu_{O_-(4)}\left(\tilde{x}\right)\, \frac{1}{P_-\left(p, \tilde{x}\right)}\, Z_{N_f}^{C^- P^-}\left(\phi, p, \tilde{x}, \tilde{y}\right) \,.
\end{align}
\end{subequations}
These branches can be used to obtain the following \emph{symmetric cases} of the Hilbert series
\begin{subequations}\label{eqn:HcasesBasic}
\begin{alignat}{1}
H^\text{tot}                     &= H^{C^+ P^+} \,, \\[8pt]
H^{C\text{-even}}                &= \frac12 \left( H^{C^+ P^+} + H^{C^- P^+} \right) \,, \\[8pt]
H^{P\text{-even}}                &= \frac12 \left( H^{C^+ P^+} + H^{C^+ P^-} \right) \,, \label{eqn:Peven} \\[8pt]
H^{C\text{-even}\,P\text{-even}} &= \frac14 \left( H^{C^+ P^+} + H^{C^+ P^-} + H^{C^- P^+} + H^{C^- P^-} \right) \,,
\end{alignat}
\end{subequations}
where we have suppressed the arguments $(\phi, p)$ and the subscript $N_f$. With the above symmetric cases, we can further derive other cases of interest, such as the various components and partial sums of the Hilbert series regarding to the $C$ and $P$ discrete symmetries, as summarized in \cref{tbl:HComponents}:
\begin{subequations}\label{eqn:HcasesMore}
\begin{alignat}{1}
H^{C\text{-even}\, P\text{-odd}}  &= H^{C\text{-even}} - H^{C\text{-even}\,P\text{-even}} \,, \\[10pt]
H^{C\text{-odd} \, P\text{-even}} &= H^{P\text{-even}} - H^{C\text{-even}\,P\text{-even}} \,, \\[10pt]
H^{C\text{-odd} \, P\text{-odd}}  &= H^{\text{tot}} - H^{C\text{-even}} - H^{P\text{-even}} + H^{C\text{-even}\,P\text{-even}} \,, \\[10pt]
H^{C\text{-odd}}                  &= H^{\text{tot}}-H^{C\text{-even}} \,, \\[10pt]
H^{P\text{-odd}}                  &= H^{\text{tot}}-H^{P\text{-even}} \,, \label{eqn:Podd} \\[10pt]
H^{CP\text{-even}}                &= H^{C\text{-even}\, P\text{-even}} + H^{C\text{-odd}\, P\text{-odd}} \,, \label{eqn:HcasesMore_CP-even} \\[10pt]
H^{CP\text{-odd}}                 &= H^{\text{tot}} - H^{CP\text{-even}} \,. \label{eqn:HcasesMore_CP-odd}
\end{alignat}
\end{subequations}
The $CP$-even and $CP$-odd series, Eqs.~\eqref{eqn:HcasesMore_CP-even} and~\eqref{eqn:HcasesMore_CP-odd}, are easily understood from the constituent relations in Eq.~\eqref{eqn:HcasesBasic}, \textit{e.g.} $H^{CP\text{-even}} = \frac{1}{2}\big(H^{C^+P^+} + H^{C^-P^-}\big)$.

\begin{table}[t]
\renewcommand{\arraystretch}{1.8}
\setlength{\arrayrulewidth}{.3mm}
\setlength{\tabcolsep}{1em}
\begin{center}
\begin{tabular}{c|cc}
$H^{\text{tot}}$    & $H^{P\text{-even}}$ & $H^{P\text{-odd}}$ \\\hline
$H^{C\text{-even}}$ & $H^{C\text{-even}\,P\text{-even}}$ & $H^{C\text{-even}\,P\text{-odd}}$ \\
$H^{C\text{-odd}}$  & $H^{C\text{-odd}\,P\text{-even}}$  & $H^{C\text{-odd}\,P\text{-odd}}$  \\
\end{tabular}\vspace{0.5cm}

$H^{CP\text{-even}} = H^{C\text{-even}\,P\text{-even}}+H^{C\text{-odd}\,P\text{-odd}}$\vspace{0.2cm}
\caption{Hilbert series split up into transformations under $C$ and $P$.}
\label{tbl:HComponents}
\end{center}
\end{table}

\section{Results}
\label{sec:results}

We begin by showing some examples of the Hilbert series we obtain using the method described in the previous section. We consider the chiral Lagrangian with two light flavours of quarks, $N_f=2$, at chiral dimension $p^6$.\footnote{We discuss the $p^4$ chiral Lagrangian in detail in \cref{appsec:p4}.} On the $C^+P^+$ branch, which counts \emph{all} operators (see \cref{eqn:HcasesBasic}), we have
\begin{align}
H^{C^+ P^+}_{N_f=2}=& D^2f_{-}^2 + 2f_{-}^3 + D^2f_{-}f_{+} + 2f_{-}^2f_{+} + D^2f_{+}^2 + 2f_{-}f_{+}^2 + 2f_{+}^3 + 2f_{-}f_{+}\Sigma_{-} + D^2\Sigma_{-}^2 \nonumber \\ &
 + 2f_{-}^2\langle\Sigma_{-}\rangle  + 2f_{-}f_{+}\langle\Sigma_{-}\rangle  + 2f_{+}^2\langle\Sigma_{-}\rangle  + \Sigma_{-}^2\langle\Sigma_{-}\rangle  + D^2\langle\Sigma_{-}\rangle^2 + \langle\Sigma_{-}\rangle^3 + 2f_{-}f_{+}\Sigma_{+} \nonumber \\ &
 + D^2\Sigma_{-}\Sigma_{+} + \Sigma_{-}\langle\Sigma_{-}\rangle \Sigma_{+} + D^2\Sigma_{+}^2 + \langle\Sigma_{-}\rangle \Sigma_{+}^2 + 2f_{-}^2\langle\Sigma_{+}\rangle  + 2f_{-}f_{+}\langle\Sigma_{+}\rangle  + 2f_{+}^2\langle\Sigma_{+}\rangle  \nonumber \\ &
 + \Sigma_{-}^2\langle\Sigma_{+}\rangle  + D^2\langle\Sigma_{-}\rangle \langle\Sigma_{+}\rangle  + \langle\Sigma_{-}\rangle^2\langle\Sigma_{+}\rangle  + \Sigma_{-}\Sigma_{+}\langle\Sigma_{+}\rangle  + \Sigma_{+}^2\langle\Sigma_{+}\rangle  + D^2\langle\Sigma_{+}\rangle^2 \nonumber \\ &
 + \langle\Sigma_{-}\rangle \langle\Sigma_{+}\rangle^2 + \langle\Sigma_{+}\rangle^3 + 2Df_{-}^2u + 4Df_{-}f_{+}u + 2Df_{+}^2u + Df_{-}\Sigma_{-}u + Df_{+}\Sigma_{-}u \nonumber \\ &
 + D\Sigma_{-}^2u + Df_{-}\langle\Sigma_{-}\rangle u + Df_{+}\langle\Sigma_{-}\rangle u + D\Sigma_{-}\langle\Sigma_{-}\rangle u + Df_{-}\Sigma_{+}u + Df_{+}\Sigma_{+}u \nonumber \\ &
 + D\Sigma_{-}\Sigma_{+}u + D\langle\Sigma_{-}\rangle \Sigma_{+}u + D\Sigma_{+}^2u + Df_{-}\langle\Sigma_{+}\rangle u + Df_{+}\langle\Sigma_{+}\rangle u + D\Sigma_{-}\langle\Sigma_{+}\rangle u \nonumber \\ &
 + D\Sigma_{+}\langle\Sigma_{+}\rangle u + D^2f_{-}u^2 + 8f_{-}^2u^2 + D^2f_{+}u^2 + 10f_{-}f_{+}u^2 + 8f_{+}^2u^2 + 2f_{-}\Sigma_{-}u^2 \nonumber \\ &
 + 2f_{+}\Sigma_{-}u^2 + 2\Sigma_{-}^2u^2 + D^2\langle\Sigma_{-}\rangle u^2 + 2f_{-}\langle\Sigma_{-}\rangle u^2 + 2f_{+}\langle\Sigma_{-}\rangle u^2 + \langle\Sigma_{-}\rangle^2u^2 \nonumber \\ &
 + 2f_{-}\Sigma_{+}u^2 + 2f_{+}\Sigma_{+}u^2 + 2\Sigma_{-}\Sigma_{+}u^2 + 2\Sigma_{+}^2u^2 + D^2\langle\Sigma_{+}\rangle u^2 + 2f_{-}\langle\Sigma_{+}\rangle u^2 \nonumber \\ &
 + 2f_{+}\langle\Sigma_{+}\rangle u^2 + \langle\Sigma_{-}\rangle \langle\Sigma_{+}\rangle u^2 + \langle\Sigma_{+}\rangle^2u^2 + 4Df_{-}u^3 + 4Df_{+}u^3 + 2D\Sigma_{-}u^3 \nonumber \\ &
 + 2D\Sigma_{+}u^3 + 2D^2u^4 + 4f_{-}u^4 + 4f_{+}u^4 + 2\langle\Sigma_{-}\rangle u^4 + 2\langle\Sigma_{+}\rangle u^4 + Du^5 + 3u^6 \,.
\label{eq:p6CPPP}
\end{align}
The above Hilbert series gives detailed information about the number of independent operators made out of the building blocks $u_\mu$ \etc appearing in \cref{tbl:Transformations}, and covariant derivatives (which have a chiral dimension of one).\footnote{See \cref{appsec:SPM} for further details on how the covariant derivative is treated in the Hilbert series formalism.} To indicate the latter we have instated a symbol $D$ as a spurion for the derivative; the power to which it appears in each term is deduced by chiral dimension counting. We emphasise that in the Hilbert series it is simply a variable, not a (differential) operator, as are all other symbols that represent fields. For example, the first term in \cref{eq:p6CPPP} represents an operator that is constructed out of two powers of $f_-$ fields, together with two covariant derivatives. The unit coefficient in front of this term indicates that there is only one independent such operator. Similarly, the second term in the above Hilbert series indicates that there are two independent operators constructed out of three $f_-$ fields, and so on.

Turning to the $C^+ P^-$ branch, where $P$-odd operators come with a negative sign, we get
\begin{align}
H^{C^+ P^-}_{N_f=2}=&D^2f_{-}^2 - D^2f_{-}f_{+} + D^2f_{+}^2 + D^2\Sigma_{-}^2 - \Sigma_{-}^2\langle\Sigma_{-}\rangle  + D^2\langle\Sigma_{-}\rangle^2  - \langle\Sigma_{-}\rangle^3 - D^2\Sigma_{-}\Sigma_{+}  \nonumber \\ &
+ \Sigma_{-}\langle\Sigma_{-}\rangle \Sigma_{+} + D^2\Sigma_{+}^2 - \langle\Sigma_{-}\rangle \Sigma_{+}^2 + \Sigma_{-}^2\langle\Sigma_{+}\rangle  - D^2\langle\Sigma_{-}\rangle \langle\Sigma_{+}\rangle  + \langle\Sigma_{-}\rangle^2\langle\Sigma_{+}\rangle   \nonumber \\ &
- \Sigma_{-}\Sigma_{+}\langle\Sigma_{+}\rangle  + \Sigma_{+}^2\langle\Sigma_{+}\rangle  + D^2\langle\Sigma_{+}\rangle^2 - \langle\Sigma_{-}\rangle \langle\Sigma_{+}\rangle^2 + \langle\Sigma_{+}\rangle^3 - Df_{-}\Sigma_{-}u  \nonumber \\ &
+ Df_{+}\Sigma_{-}u - D\Sigma_{-}^2u - Df_{-}\langle\Sigma_{-}\rangle u + Df_{+}\langle\Sigma_{-}\rangle u - D\Sigma_{-}\langle\Sigma_{-}\rangle u + Df_{-}\Sigma_{+}u  \nonumber \\ &
- Df_{+}\Sigma_{+}u + D\Sigma_{-}\Sigma_{+}u + D\langle\Sigma_{-}\rangle \Sigma_{+}u - D\Sigma_{+}^2u + Df_{-}\langle\Sigma_{+}\rangle u - Df_{+}\langle\Sigma_{+}\rangle u  \nonumber \\ &
+ D\Sigma_{-}\langle\Sigma_{+}\rangle u - D\Sigma_{+}\langle\Sigma_{+}\rangle u - D^2f_{-}u^2 + 2f_{-}^2u^2 + D^2f_{+}u^2 + 2f_{+}^2u^2 + 2\Sigma_{-}^2u^2  \nonumber \\ &
- D^2\langle\Sigma_{-}\rangle u^2 + \langle\Sigma_{-}\rangle^2u^2 - 2\Sigma_{-}\Sigma_{+}u^2 + 2\Sigma_{+}^2u^2 + D^2\langle\Sigma_{+}\rangle u^2 - \langle\Sigma_{-}\rangle \langle\Sigma_{+}\rangle u^2  \nonumber \\ &
+ \langle\Sigma_{+}\rangle^2u^2 + 2Df_{-}u^3 - 2Df_{+}u^3 + 2D\Sigma_{-}u^3 - 2D\Sigma_{+}u^3 + 2D^2u^4 - 2\langle\Sigma_{-}\rangle u^4  \nonumber \\ &
+ 2\langle\Sigma_{+}\rangle u^4 - Du^5 + 3u^6 \,.
\label{eq:p6CPPM}
\end{align}
Note that the Hilbert series $H^{C^+ P^-}$, $H^{C^- P^+}$, and $H^{C^- P^-}$ generically contain negative terms such that once combined with $H^{C^+ P^+}$ as in \cref{eqn:HcasesBasic}, they make the Hilbert series that count operators of definite symmetry, where all terms will be positive, and indeed integer.

As a simple check, one can readily verify that \cref{eq:p6CPPP,eq:p6CPPM} can be combined as per \cref{eqn:Peven,eqn:Podd} to produce parity even and odd Hilbert series that  only contain terms with positive, integer coefficients. Note how, for example, the penultimate terms in \cref{eq:p6CPPP,eq:p6CPPM}, $\pm D u^5$ (five $u_\mu$ fields and one derivative, which is parity odd in the case $N_f=2$), cancel each other in the sum \cref{eqn:Peven} to produce the $P$-even Hilbert series.

As mentioned in the introduction, it is common in the literature to separate out operators which include a spacetime epsilon tensor $\epsilon^{\mu\nu\rho\sigma}$---denoting these `anomalous' terms, for example see \cite{Bijnens:2001bb} at chiral dimension $p^6$. This information is also available with our method, using the fact that the $\epsilon$ tensor changes sign under parity transformations. For overall $P$-even operators, such epsilon terms must have an odd number of intrinsic parity odd fields. Writing the dependence on variables explicitly, one can define a flipped version of the Hilbert series
\begin{equation}
H_\text{flipped}(u,\Sigma_+,\Sigma_-,\langle\Sigma_+\rangle,\langle\Sigma_-\rangle,f_+,f_-)
= H(-u,\Sigma_+,-\Sigma_-,\langle\Sigma_+\rangle,-\langle\Sigma_-\rangle,f_+,-f_-) \,,
\end{equation}
\ie variables corresponding to fields with negative intrinsic parity are negated, such that the Hilbert series without `anomalous' operators is given by
\begin{equation}
H_{\text{no-}\epsilon}^{P\text{-even}} = \frac12 \left(H^{P\text{-even}} + H_\text{flipped}^{P\text{-even}}\right) \,.
\label{eq:Hflip}
\end{equation}
It follows that a Hilbert series for the anomalous terms only is constructed as
\begin{equation}
H_{\epsilon \text{-only}} = H - H_{\text{no-}\epsilon} \,.
\end{equation}
For $P$-odd Hilbert series, the above logic is reversed---epsilon terms must have an even number of intrinsic parity odd fields, and the plus sign in \cref{eq:Hflip} gets replaced by a minus sign.

\begin{table}
\renewcommand{\arraystretch}{1.5}
\begin{center}
{\small
\begin{tabular}{c lllllll}
chiral dim & $SU(2)$ & $SU(3)$ & $SU(4)$ & $SU(5)$ & $SU(6)$ & $SU(7)$ & $SU(8)$ \\
\hline
$p^{2}  $ & 2 (0)   $\rightarrow$      &        &     & &  &  & \\
$p^{4}  $ & 10 (0)       & 12 (0)       & 13 (0)  $\rightarrow$  & & & &  \\
$p^{6}  $ & 56 (5)      & 94 (23)        &112 (24)        &114 (24)&115 (24) $\rightarrow$&&  \\
$p^{8}  $ & 475 (92)     & 1254 (705)    & 1752 (950)& 1839 (998) & 1859 (999) & 1861 (999) & 1862 (999)  \\
\end{tabular}
}
\caption{Enumeration of \emph{both} $C$-even \emph{and} $P$-even operators in the chiral Lagrangian with $2\le N_f\le 8 $ light quarks ($SU(N_f)$ unbroken symmetry). Numbers not in parentheses count non-anomalous operators (\ie excluding operators which involve an $\epsilon^{\mu\nu\rho\sigma}$), while numbers in  parentheses count only the anomalous operators. The arrow  `$\rightarrow$' denotes that the entry is repeated to complete the row.}
\label{tab:suncount}
\end{center}
\end{table}

\begin{table}
\renewcommand{\arraystretch}{1.5}
\begin{center}
{\small
\begin{tabular}{ c  c  c  c c }
\multicolumn{5}{c}{$N_f=2$} \\
chiral dim & Total&   $C$-even & $P$-even & $CP$-even  \\
\hline
$p^{6}  $ & 151    & 103 & 82& 88  \\
$p^{8}  $ & 1834   & 1050& 943 & 975 \\
\end{tabular}\hspace{10pt}
\begin{tabular}{ c   c  c  c }
\multicolumn{4}{c}{$N_f=3$} \\
Total&   $C$-even & $P$-even & $CP$-even  \\
\hline
315    & 206 & 165 & 178  \\
6882 &   3768&  3479 &  3553  \\
\end{tabular}
}
\caption{Enumeration of operators in the chiral Lagrangian broken down by behaviour under $C$, $P$, and $CP$ transformations, for the cases of $N_f=2$  and $N_f=3$ light quark flavours. In contrast to Table~\ref{tab:suncount}, `anomalous' terms which involve an $\epsilon^{\mu\nu\rho\sigma}$ are not separated out, and are included in the enumeration. Note that $CP$-even counts both $C$-even $P$-even and $C$-odd $P$-odd operators.}
\label{tab:cpbreakdown}
\end{center}
\end{table}

Of course, one can ``coarse-grain'' the information; setting all of the variables $u$, $\Sigma_\pm$, $\langle\Sigma_\pm\rangle$, $f_\pm$, and $D$ to unity in a Hilbert series, one obtains the total number of independent operators. We will present a few results using this coarse graining, but we stress that Hilbert series with full field content information---as in \cref{eq:p6CPPP}---have greater utility than simply providing an overall enumeration and indeed contain information much more useful for the actual \emph{construction} of operators (see~\cite{Henning:2017fpj}, and developments \eg~\cite{Henning:2019enq,Henning:2019mcv}).

\cref{tab:suncount} summarises the coarse-grained Hilbert series output for the both $C$-even and $P$-even chiral Lagrangian with $2\le N_f\le 8$ flavours, at chiral dimension $p^2$ through $p^8$, providing the enumeration of both non-anomalous  and anomalous operators, with the latter being the number given in parentheses. We find agreement with the most up to date results in the literature (accounting for missed relations as summarised in~\cite{Bijnens:2018lez}). Concretely, the known results are the non-anomalous operators at chiral dimension $p^4$~\cite{Gasser:1983yg,Gasser:1984gg}, $p^6$~\cite{Fearing:1994ga,Bijnens:1999sh,Haefeli:2007ty} and $p^8$~\cite{Bijnens:2018lez}, and the anomalous operators at chiral dimension $p^4$ (of which there are none, see e.g.~\cite{Wess:1971yu}) and $p^6$~\cite{Ebertshauser:2001nj,Bijnens:2001bb}, for the physical cases $SU(2)$, $SU(3)$, and for the asymptotic number in each row, which corresponds to what is denoted $SU(N_f)$ in the literature.\footnote{Regarding the agreement at order $p^2$ and $p^4$, we point the reader to the comments made in \cref{subsec:IBP}.}\,\footnote{The `general $N_f$' flavours counting, or `$SU(N_f)$ case', is often presented in the literature, meaning no $SU(N_f)$ group theory relations are imposed to reduce the number of operators. More concretely, at a given chiral dimension $p^k$, this number can be taken to mean the $SU(N_f \ge k)$ counting, which corresponds to the asymptotic numbers in each row of \cref{tab:suncount}. These numbers are actually technically more difficult to obtain with the Hilbert series method (on account of the more complicated characters/group integral) than the physical cases.}

The main new result shown in \cref{tab:suncount} is the enumeration of  $C$-even {\it and} $P$-even anomalous operators at chiral dimension $p^8$, thus completing the enumeration all $C$-even {\it and} $P$-even operators at this order---we present a more detailed breakdown in \cref{appsec:p8}. Also new are the enumeration of operators at $p^k$ in the (non-physical) cases where the number of light quarks $3< N_f<k$.

\begin{figure}
\begin{center}
\includegraphics[width=12cm]{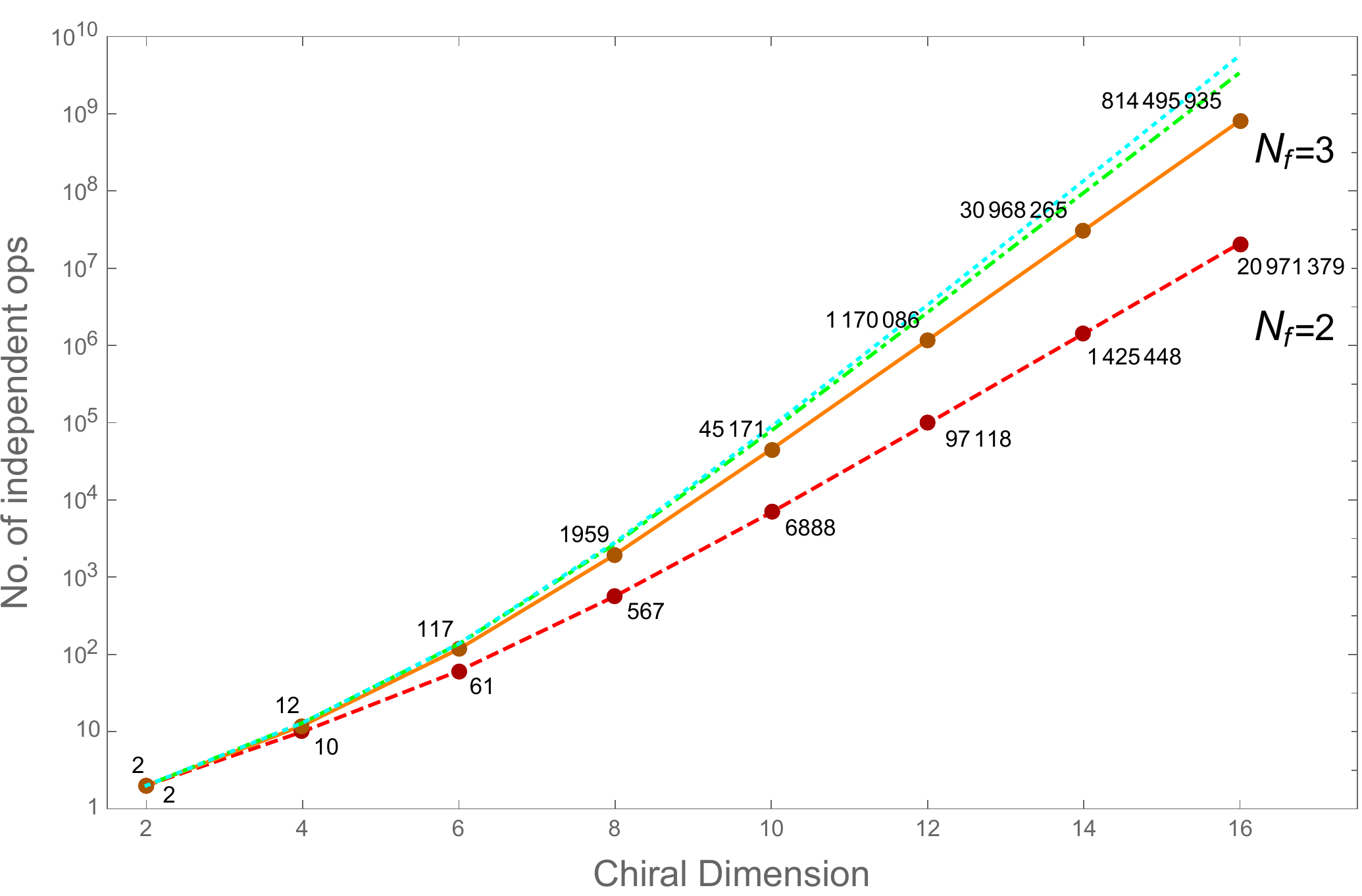}
\caption{The number of independent operators in the $C$-even $P$-even chiral Lagrangian as a function of chiral dimension, up to $p^{16}$. Red dashed line, through points numbered 2, 10, 61, $\ldots$, corresponds to all operators  in the case of two light quark flavours, $N_f=2$.  Orange solid line, through points numbered 2, 12, 117, $\ldots$, corresponds to all operators  in the case of three light quark flavours, $N_f=3$. The green dot-dashed and cyan dotted lines (without numbered dots) are, respectively, the cases $N_f=4$ and $N_f=5$ (the enumeration is provided in Appendix~\ref{appsec:nf45}).}
\label{fig:growth}
\end{center}
\end{figure}

In \cref{tab:cpbreakdown}, we show the number of $C$-even, $P$-even, and $CP$-even ({\it i.e.} including both  $C$-even $P$-even and $C$-odd $P$-odd) operators at chiral dimension $p^6$ and $p^8$, for the physically relevant cases $N_f=2, 3$. The number of $P$-even and $P$-odd operators are roughly equal, as might be expected from the fact that there are two versions of most fields, one with even intrinsic parity, and one with odd. On the other hand, we observe there are somewhat more $C$-even operators than $C$-odd at these chiral dimensions. Furthermore, comparing the entries in \cref{tab:cpbreakdown} to those in  \cref{tab:suncount}, we see that the number of $C$-odd {\it and} $P$-odd (and hence $CP$-even) operators is only roughly $30-40\%$ of the number of $C$-even {\it and} $P$-even at chiral dimension $p^6$, and $70-80\%$ at chiral dimension $p^8$. Further results on $C$-odd {\it and} $P$-odd operators, as well as $CP$-odd operators, can be found in the accompanying {\tt Mathematica} notebook. All  results  shown in \cref{tab:cpbreakdown} are, to the best of our knowledge, new.

Finally, in \cref{fig:growth} we look at the growth of $C$-even {\it and} $P$-even operators for $N_f=2, 3$ as the chiral dimension grows large, up to $p^{16}$. As expected on general grounds (see \eg the discussion in~\cite{Henning:2017fpj}) the number of independent operators grows exponentially. Similar growth was observed in the SM EFT~\cite{Henning:2015alf}; for the mesonic QCD chiral Lagrangian we see that the growth of operators is smoother as all building blocks are bosonic, so the variations evident in moving between even and odd mass dimensions in the SM EFT are not present. We also plot curves which show the growth of operators for the unphysical cases of $N_f=4,5$ for comparison (with  enumeration  provided in Appendix~\ref{appsec:nf45}); at fixed chiral dimension we observe the number of operators converging to a fixed value with increasing $N_f$, as seen in the rows of \cref{tab:suncount}.

\section{Discussion}

In summary, we have adapted the Hilbert series technology so as to apply it to the enumeration of operators in the mesonic QCD chiral Lagrangian. This provides a systematic way to determine operator content at a given chiral dimension. We confirmed existing results in the literature; new results presented include the $C$-even \emph{and} $P$-even operator content of the anomalous chiral Lagrangian at chiral dimension $p^8$, and  the $C$-even, $P$-even, and $CP$-even operator content at chiral dimension $p^6$ and $p^8$. We augmented aspects of the Hilbert series method, most notably through the inclusion of the operation of charge conjugation via the folding of $\frak{su}(n)$ Dynkin diagrams, as well as previously unconsidered field content.

We conclude here with a discussion of an interesting possible application of our work concerning the rare decays of hadrons. This is inspired by the recent results from the KOTO experiment at J-PARC, which reported possible excess events in $K_L\rightarrow \pi^0 \nu \bar{\nu}$ \cite{Shinohara:2020brf}. If taken literally, it appears to violate the well-known Grossman-Nir bound \cite{Grossman:1997sk}. The bound is based on the assumptions of isospin and lepton-flavor conservation, which
forces the decay to be a $CP$-violating effect at the leading order in the EFT.
Yet they pointed out that higher-order $CP$-conserving operators can contribute to the process. In addition, isospin violation and/or lepton-flavor violation also open up possible loopholes. We believe our classification of higher-order operators in the chiral Lagrangian facilitates the study of identifying possible sources of higher-order operators with new flavor violations. Even though higher-order operators are suppressed when the new physics scale is above the electroweak scale, this is a place where the Standard Model contribution is so suppressed that they can play an important role.  In addition, there are models with light new particles ({\it e.g.}\/, \cite{Egana-Ugrinovic:2019wzj,Dev:2019hho,Jho:2020jsa,Liu:2020qgx,Dutta:2020scq,Liu:2020ser,Ziegler:2020ize,He:2020jzn,He:2020jly}). In this case, our classification can be straightforwardly expanded to include new light degrees of freedom in the Hilbert series.

\acknowledgments
H.M. thanks Hiraku Nakajima for useful discussions.
B.H. is supported by the Swiss National Science Foundation, under grant no. PP00P2-170578.
X.L. is supported by the U.S. Department of Energy, under grant number DE-SC0011640.
T.M. is supported by the World Premier International Research Center Initiative (WPI) MEXT, Japan, and by JSPS KAKENHI grants JP18K13533, JP19H05810, JP20H01896, and JP20H00153.
The work of HM was supported by the NSF grant PHY-1915314, by the U.S. DOE Contract DE-AC02-05CH11231, by the JSPS Grant-in-Aid for Scientific Research JP17K05409, MEXT Grant-in-Aid for Scientific Research on Innovative Areas JP15H05887, JP15K21733, by WPI, MEXT, Japan, and Hamamatsu Photonics, K.K.

\appendix
\section{Characters of single particle modules}
\label{appsec:SPM}

Schematically, the single particle modules relevant for the chiral Lagrangian are
\begin{equation}\renewcommand\arraystretch{1.5}
R_u^{} = \mqty( u_\mu \\ D_{\mu_1} u_\mu \\ D_{\mu_1} D_{\mu_2} u_\mu \\ \vdots )
\quad,\quad
R_\Sigma^{} = \mqty( \Sigma \\ D_{\mu_1} \Sigma \\ D_{\mu_1} D_{\mu_2} \Sigma \\ \vdots )
\quad,\quad
R_f^{} = \mqty( f_{\mu\nu} \\ D_{\mu_1} f_{\mu\nu} \\ D_{\mu_1} D_{\mu_2} f_{\mu\nu} \\ \vdots ) \,\,.
\end{equation}
Here we use $\Sigma$ to cover the cases $\Sigma_\pm, \langle\Sigma_\pm\rangle$, and $f$ to cover the cases $f_\pm$. We describe the above list of components as being `schematic', because many components are actually absent/vanishing due to additional properties satisfied by the single particle modules. There are three of these properties:
\begin{subequations}
\begin{alignat}{2}
\text{Equations of Motion}:&\quad  & D^\mu u_\mu &= 0 \,, \\[8pt]
\text{Lie Algebra Relations}:&\quad  & D_\mu u_\nu - D_\nu u_\mu &= 0 \,, \\[8pt]
\text{Bianchi Identities}:&\quad  & D_\rho f_{\mu\nu} + D_\mu f_{\nu\rho} + D_\nu f_{\rho\mu} &= 0 \,.
\end{alignat}
\end{subequations}
Precisely speaking, the listed properties do not make these three combinations zero, but actually obtainable from components with less number of derivatives (see \eg \cite{Bijnens:1999sh,Bijnens:2018lez}). Therefore, they can be treated as zero in computing the Hilbert series. With the same spirit, the covariant derivatives can be treated as commuting objects: $\comm{D_\mu}{D_\nu}=0$.

With the above, we can find the $SO(4)$ representations for all the components in the single particle modules
\begin{subequations}
\begin{align}
D^n u_\mu :&\quad  (n+1, 0)  \,, \\[8pt]
D^n \Sigma :&\quad  \text{sym}^n(1, 0)  \,, \\[5pt]
D^n f_{\mu\nu} :&\quad  \text{sym}^n(1, 0)\otimes \Big((1, 1)\oplus(1, -1)\Big) - \text{sym}^{n-1}(1, 0)\otimes(1, 0) + \text{sym}^{n-2}(1, 0)  \,.
\end{align}
\end{subequations}
In the last line above, the `$-$' sign of the second term should be understood as a quotient; it makes sense because the second term is a subspace of the first term. From these representations, it is straightforward to compute the explicit expressions of the characters $\chi_i^{P^\pm}$ for all the single particle modules. The results are
\begin{subequations}\label{eqn:chilistSpacetime}
\begin{align}
\chi_u^{P^+} \left(p, x\right) &= \left(1 - p^2\right) P_+\left(p, x_1, x_2 \right) - 1 \,, \\[10pt]
\chi_u^{P^-} \left(p, \tilde{x}\right) &= -p\left(1 - p^2\right)\left( x_1 + x_1^{-1} - p \right) P_-\left(p, x_1\right) \,, \\[10pt]
\chi_{\Sigma_\pm}^{P^+} \left(p, x\right) = \chi_{\langle\Sigma_\pm\rangle}^{P^+} \left(p, x\right) &= p^2 P_+\left(p, x_1, x_2 \right) \,, \\[10pt]
\chi_{\Sigma_\pm}^{P^-} \left(p, \tilde{x}\right) = \chi_{\langle\Sigma_\pm\rangle}^{P^-} \left(p, \tilde{x}\right) &= \pm p^2 P_-\left(p, x_1\right) \,, \\[10pt]
\chi_{f_\pm}^{P^+} \left(p, x\right) &= p^2 \bigg[ x_1x_2 + 1 + \frac{1}{x_1x_2} + \frac{x_1}{x_2} + 1 + \frac{x_2}{x_1} \notag\\
&\hspace{40pt} - p \left(x_1+x_1^{-1}+x_2+x_2^{-1}\right) + p^2 \bigg] P_+\left(p, x_1, x_2 \right) \,, \\[10pt]
\chi_{f_\pm}^{P^-} \left(p, \tilde{x}\right) &= \pm p^3\left(x_1+x_1^{-1}-p\right) P_-\left(p, x_1\right) \,,
\end{align}
\end{subequations}
with the definitions
\begin{subequations}
\begin{align}
P_+\left(p, x_1, x_2 \right) &= \frac{1}{\left(1-p x_1\right) \left(1-p x_1^{-1}\right) \left(1-p x_2\right) \left(1-p x_2^{-1}\right)} \,, \\[8pt]
P_-\left(p, x_1\right) &= \frac{1}{\left(1 - p x_1\right) \left(1 - p x_1^{-1}\right) \left(1- p^2\right)} \,.
\end{align}
\end{subequations}
For completeness, we also provide the explicit expressions of the characters $\chi_{i,\,N_f}^{C^\pm}$ for all the single particle modules
\begin{subequations}\label{eqn:chilistInternal}
\begin{align}
\chi_{\langle\Sigma_\pm\rangle,\,N_f}^{C^+} \left(y\right) = \chi_{\langle\Sigma_\pm\rangle,\,N_f}^{C^-} \left(\tilde{y}\right) &= 1 \,, \\[10pt]
\chi_{u,\,N_f}^{C^+} (y) = \chi_{\Sigma_\pm,\,N_f}^{C^+} (y) = \chi_{f_\pm,\,N_f}^{C^+} (y) &= \chi_\text{adjoint}^{SU(N_f)}(y) \,, \\[10pt]
\chi_{u,\,N_f=2k}^{C^-} \left(\tilde{y}\right) = \chi_{\Sigma_\pm,\,N_f=2k}^{C^-} \left(\tilde{y}\right) = \mp \chi_{f_\pm,\,N_f=2k}^{C^-} \left(\tilde{y}\right) &= -\chi_\text{fundamental}^{SO(2k+1)}\left(\tilde{y}\right) \,, \\[10pt]
\chi_{u,\,N_f=2k+1}^{C^-} \left(\tilde{y}\right) = \chi_{\Sigma_\pm,\,N_f=2k+1}^{C^-} \left(\tilde{y}\right) = \mp \chi_{f_\pm,\,N_f=2k+1}^{C^-} \left(\tilde{y}\right) &= -\chi_\text{fundamental}^{Sp(2k)}\left(\tilde{y}\right) \,.
\end{align}
\end{subequations}

\section{Folding for charge conjugation}
\label{appsec:folding}

In order to impose charge conjugation invariance via Hilbert series, we need to figure out the characters (and Haar measure) on the odd branch of the orbit group $\widetilde{SU}(N_f)\equiv SU(N_f)\rtimes\mathcal{C}$. These are summarized in the main text (\cref{tbl:chiBranchesGN}) for the representations relevant to the chiral Lagrangian. In this appendix, we show how to derive these results.

Consider an arbitrary irrep of $SU(N_f)$. If it does not form a $\widetilde{SU}(N_f)$ rep by itself, one needs to find its charge conjugation partner rep, and pair them up to form an irrep of $\widetilde{SU}(N_f)$. In such $\widetilde{SU}(N_f)$ irreps, the group elements on the odd branch $g_-\in \widetilde{SU}_-(N_f)$ are off-block-diagonal and hence have vanishing characters, $\chi_-=\tr\left(g_-\right)=0$. The more nontrivial case is that the given $SU(N_f)$ irrep is self-conjugate under $C$ and hence forms a $\widetilde{SU}(N_f)$ irrep by itself.\footnote{In fact, each such self-conjugate $SU(N_f)$ irrep can form two distinct $\widetilde{SU}(N_f)$ irreps, depending on an intrinsic sign choice $\eta_C^{}=\pm$ in its transformation under $C$. Consequently, there is an overall sign in the character for the odd branch elements, as reflected in \cref{tbl:chiBranchesGN}.} In this case, $\chi_-$ follows from the $C$-invariant weights of the irrep, which are obtained from the ($C$-invariant) highest weight by subtracting $C$-invariant linear combinations of the simple roots of $SU(N_f)$. These invariant combinations are in turn generated by a new set of simple roots, which can be obtained by folding the Dynkin diagram $A_r$ (with $r=N_f-1$) representing the Lie Algebra $\frak{su}(r+1)$. In fact, there are two kinds of folding that one can define: folding by \emph{average} and folding by \emph{sum}. The former gives us the $C$-invariant subalgebra; and the latter gives us the $C$-invariant weight lattice, which is what we need in this appendix. (See \cite{saito1985extended} and also App. C.2 in \cite{Henning:2017fpj} for details.) In what follows, we will show that $A_{2k-1}$ folded by sum yields $B_k$, corresponding to the root system of $\frak{so}(2k+1)$; $A_{2k}$ folded by sum yields $C_k$, corresponding to the root system of $\frak{sp}(2k)$. The results in \cref{tbl:chiBranchesGN} in the main text hence follow.

\subsection{Root and weight systems}

We first summarize the root and weight system for $A_{r}=\frak{su}(r+1)$, as well as its orbit groups $B_k=\frak{so}(2k+1)$ and $C_k=\frak{sp}(2k)$. The roots are vectors on the root lattice generated by simple roots $r_{i}$ obtained from a Cartan matrix $A$
\begin{align}
	A_{ij} = 2 \frac{r_{i} \cdot r_{j}}{r_{i} \cdot r_{i}} \,.
\end{align}
The diagonal elements of Cartan matrix are all $A_{ii}=2$, while non-diagonal elements are non-positive $A_{ij} \leq 2$.  It is required that $A=DS$ where $D$ is a diagonal matrix while $S$ is symmetric.  This requirement allows for a classification of Cartan matrices.  Dynkin diagrams are graphical representation of Cartan matrices.  The weights are vectors on the weight lattice generated by fundamental weights $w_{i}$ defined by the simple roots
\begin{align}
	2 \frac{w_{i} \cdot r_{j}}{r_{j} \cdot r_{j}} = \delta_{ij} \,.
\end{align}

For $A_{r}=\frak{su}(r+1)$, the Cartan matrix has $A_{i\, i}=2$, $A_{i\, i+1}=A_{i+1\, i}=-1$ for $i=1, \cdots, r-1$ and otherwise zero,
\begin{align}
	A =
	\left( \begin{array}{rrrcrrr}
		2 & -1 &  0& \cdots & 0 & 0 & 0 \\
		-1 & 2 & -1 & \cdots & 0 & 0 & 0 \\
		0 & -1 & 2 & \cdots & 0 & 0 & 0 \\
		\vdots & \vdots & \vdots & \ddots & \vdots & \vdots & \vdots \\
		0 & 0 & 0 & \cdots & 2 & -1 & 0\\
		0 & 0 & 0 & \cdots & -1 & 2 & -1\\
		0 & 0 & 0 & \cdots & 0 & -1 & 2
		\end{array} \right) \,.
\end{align}
It is convenient to use $(r+1)$-dimensional vector space, where all roots are orthogonal to the vector $\underbrace{(1,1,\cdots, 1)}_{r+1}$. The simple roots are
\begin{equation}\renewcommand\arraystretch{1.3}
\begin{array}{rrrrrrrrl}
  \alpha_1     =& (1,& -1,&  0,& \cdots,& 0,&  0,&  0)& \\
  \alpha_2     =& (0,&  1,& -1,& \cdots,& 0,&  0,&  0)& \\
                & \vdots &&&&&&& \\
  \alpha_{r-1} =& (0,&  0,&  0,& \cdots,& 1,& -1,&  0)& \\
  \alpha_{r}   =& (0,&  0,&  0,& \cdots,& 0,&  1,& -1)&
\end{array}\,\,.
\label{eqn:rtSUN}
\end{equation}
The complete set of roots is given by
\begin{equation}
  (\cdots, \pm 1, \cdots, \mp 1, \cdots) \,.
\end{equation}
There are $r(r+1)$ of them. Together with the $r$ Cartan generators, they form the set of $(r+1)^2-1$ generators. The fundamental weights are
\begin{equation}\renewcommand\arraystretch{2.0}
\begin{array}{rrrrrrrrl}
  \mu_1     =& \dfrac12\, (1,& -1,& -1,& \cdots,& -1,& -1,& -1)& \\
  \mu_2     =& \dfrac12\, (1,&  1,& -1,& \cdots,& -1,& -1,& -1)& \\
             & \vdots &&&&&&& \\
  \mu_{r-1} =& \dfrac12\, (1,&  1,&  1,& \cdots,&  1,& -1,& -1)& \\
  \mu_{r}   =& \dfrac12\, (1,&  1,&  1,& \cdots,&  1,&  1,& -1)&
\end{array}\,\,.
\label{eqn:fwSUN}
\end{equation}
The (first) fundamental representation has its highest weight as the first fundamental weight $\mu_1$, and all the other weights further obtained from it:
\begin{equation}
  \frac12\, (-1, \cdots, -1, +1, -1, \cdots, -1) \,.
\end{equation}
There are in total $r+1$ of them, including the highest one.

For $B_{k}=\frak{so}(2k+1)$, the Cartan matrix has $A_{i\, i}=2$, $A_{i\, i+1}=A_{i+1\, i}=-1$ for $i=1, \cdots, k-2$, $A_{k-1\, k}=-1$, while $A_{k\, k-1}=-2$ and otherwise zero,
\begin{align}
	A =
	\left( \begin{array}{rrrcrrr}
		2 & -1 &  0& \cdots & 0 & 0 & 0 \\
		-1 & 2 & -1 & \cdots & 0 & 0 & 0 \\
		0 & -1 & 2 & \cdots & 0 & 0 & 0 \\
		\vdots & \vdots & \vdots & \ddots & \vdots & \vdots & \vdots \\
		0 & 0 & 0 & \cdots & 2 & -1 & 0\\
		0 & 0 & 0 & \cdots & -1 & 2 & -1\\
		0 & 0 & 0 & \cdots & 0 & -2 & 2
		\end{array} \right) \,.
\end{align}
The simple roots are
\begin{equation}\renewcommand\arraystretch{1.3}
\begin{array}{rrrrrrrr}
\beta_1     =& (1,& -1,&  0,& 0,& \cdots,& 0)& \\
\beta_2     =& (0,&  1,& -1,& 0,& \cdots,& 0)& \\
             & \vdots  &&&&&&\\
\beta_{k-1} =& (0,&  0,& \cdots,& 0,& 1,& -1)& \\
\beta_k     =& (0,&  0,& \cdots,& 0,& 0,&  1)&
\end{array}\,\,.
\label{eqn:rtSO2k1}
\end{equation}
Note that the last one $\beta_k$ is a \emph{short} root. The fundamental weights of $\frak{so}(2k+1)$ are
\begin{equation}\renewcommand\arraystretch{2.0}
\begin{array}{rrrrrrrr}
\nu_1     =&          (1,&  0,&  0,& \cdots,& 0,& 0)& \\
\nu_2     =&          (1,&  1,&  0,& \cdots,& 0,& 0)& \\
           &          \vdots  &&&&&&\\
\nu_{k-1} =&          (1,&  1,&  1,& \cdots,& 1,& 0)& \\
\nu_k     =& \dfrac12 (1,&  1,&  1,& \cdots,& 1,& 1)&
\end{array}\,\,.
\label{eqn:fwSO2k1}
\end{equation}

For $C_{k}=\frak{sp}(2k)$, the Cartan matrix has $A_{i\, i}=2$, $A_{i\, i+1}=A_{i+1\, i}=-1$ for $i=1, \cdots, k-2$, $A_{k\, k-1}=-1$, while $A_{k-1\, k}=-2$ and otherwise zero,
\begin{align}
	A =
	\left( \begin{array}{rrrcrrr}
		2 & -1 &  0& \cdots & 0 & 0 & 0 \\
		-1 & 2 & -1 & \cdots & 0 & 0 & 0 \\
		0 & -1 & 2 & \cdots & 0 & 0 & 0 \\
		\vdots & \vdots & \vdots & \ddots & \vdots & \vdots & \vdots \\
		0 & 0 & 0 & \cdots & 2 & -1 & 0\\
		0 & 0 & 0 & \cdots & -1 & 2 & -2\\
		0 & 0 & 0 & \cdots & 0 & -1 & 2
		\end{array} \right) \,.
\end{align}
The simple roots are
\begin{equation}\renewcommand\arraystretch{1.3}
\begin{array}{rlrrrrrr}
\gamma_1     =& (1,& -1,&  0,& 0,& \cdots,& 0)& \\
\gamma_2     =& (0,&  1,& -1,& 0,& \cdots,& 0)& \\
             & \vdots  &&&&&&\\
\gamma_{k-1} =& (0,&  0,& \cdots,& 0,& 1,& -1)& \\
\gamma_k     =& (0,&  0,& \cdots,& 0,& 0,&  2)&
\end{array}\,\,.
\label{eqn:rtSp2k}
\end{equation}
Note that the last one $\gamma_k$ is a \emph{long} root. The fundamental weights of $\frak{sp}(2k)$ are
\begin{equation}\renewcommand\arraystretch{1.3}
\begin{array}{rrrrrrrr}
\rho_1     =& (1,&  0,&  0,& \cdots,& 0,& 0)& \\
\rho_2     =& (1,&  1,&  0,& \cdots,& 0,& 0)& \\
            & \vdots  &&&&&&\\
\rho_{k-1} =& (1,&  1,&  1,& \cdots,& 1,& 0)& \\
\rho_k     =& (1,&  1,&  1,& \cdots,& 1,& 1)&
\end{array}\,\,.
\label{eqn:fwSp2k}
\end{equation}

\subsection{Folding $A_{2k-1}$}

\begin{figure}[t]
\centering
\subfigure{
\begin{tikzpicture}
	\draw (0,0) -- (0.6,0);
	\draw (1.4, 0) -- (4.6,0);
	\draw (5.4, 0) -- (6,0);	
	\draw[fill=white] (0,0) circle(.1);
	\draw[fill=white] (2,0) circle(.1);
	\draw[fill=white] (3,0) circle(.1);
	\draw[fill=white] (4,0) circle(.1);
	\draw[fill=white] (6,0) circle(.1);	
	\node at (1,0) {$\cdots$};
	\node at (5,0) {$\cdots$};
	\node at (0,0.6) {$1$};	
	\node at (2,0.6) {$k-1$};	
	\node at (3,0.6) {$k$};	
	\node at (6,0.6) {$2k-1$};	
	\node at (-1.2,0) {$A_{2k-1}$};	
\end{tikzpicture}
}\\[10pt]
\subfigure{
\begin{tikzpicture}
	\draw (0,0) -- (0.7,0);
	\draw (1.3, 0) -- (2,0);
	\draw (2,0.1) -- (3,0.1);
	\draw (2,-0.1) -- (3,-0.1);
	\draw (2.4,0.2) -- (2.6,0);
	\draw (2.4,-0.2) -- (2.6,0);		
	\draw[fill=white] (0,0) circle(.1);
	\draw[fill=white] (2,0) circle(.1);
	\draw[fill=white] (3,0) circle(.1);
	\draw[fill=white] (6.2,0) circle(.0);	
	\node at (1,0) {$\cdots$};
	\node at (-1,0) {$B_k$};	
\end{tikzpicture}
}
\caption{Folding the Dynkin diagram $A_{2k-1}$ by sum yields the Dynkin diagram $B_k$.}
\label{fig:A2km1folding}
\end{figure}
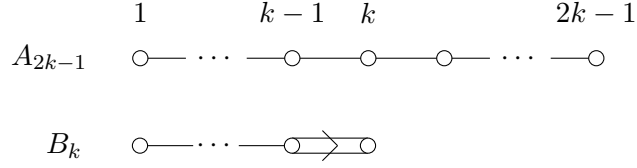

Let us first discuss the case $r=2k-1$, an odd number. In this case, there is a middle node in the Dynkin diagram---the simple root $\alpha_k$. The folding is defined by adding columns of the Cartan matrix transformed by the automorphism.  Note that the middle node is invariant under the automorphism and there is no sum. In terms of roots, it corresponds to $\alpha_{i} \rightarrow \alpha_{i} + \alpha_{2k-i}$ except for $\alpha_{k} \rightarrow \alpha_{k}$. For example in the case of $A_{5}$, the folding is
\begin{align}
	\left( \begin{array}{rrrrr}
		2 & -1 & 0 & 0 & 0\\
		-1 & 2 & -1 & 0 & 0\\
		0 & -1 & 2 & -1 & 0\\
		0 & 0 & -1 & 2 & -1\\
		0 & 0 & 0 & -1 & 2
		\end{array} \right)
	\rightarrow
	\left( \begin{array}{rrr}
		2 & -1 & 0 \\
		-1 & 2 & -1 \\
		0 & -2 & 2 \\
		-1 & 2 & -1 \\
		2 & -1 & 0
		\end{array} \right) \,.
\end{align}
The last two rows are clearly redundant.  Removing them, we obtain the Cartan matrix of $B_3$. The procedure is depicted by \cref{fig:A2km1folding}. The folding yields the following new simple roots:
\begin{equation}\renewcommand\arraystretch{1.5}
\begin{array}{rlrrrrrrrrrrrr}
\tilde\beta_1     =& \alpha_1 + \alpha_{2k-1}    &=& (1,& -1,&  0,& 0,& \cdots,& \cdots,& 0,& 0,&  1,& -1)& \\
\tilde\beta_2     =& \alpha_2 + \alpha_{2k-2}    &=& (0,&  1,& -1,& 0,& \cdots,& \cdots,& 0,& 1,& -1,&  0)& \\
                   &\vdots                       & & \vdots &&&&&&&&&&\\
\tilde\beta_{k-1} =& \alpha_{k-1} + \alpha_{k+1} &=& (0,& \cdots,& 0,& 1,& -1,&  1,& -1,& 0,& \cdots,& 0)& \\
\tilde\beta_k     =& \alpha_k                    &=& (0,& \cdots,& 0,& 0,&  1,& -1,&  0,& 0,& \cdots,& 0)&
\end{array}\,\,.
\label{eqn:tildebeta}
\end{equation}
In terms of a root system, these are equivalent to the following set
\begin{equation}\renewcommand\arraystretch{1.3}
\begin{array}{rrrrrrrr}
\beta_1     =& (1,& -1,&  0,& 0,& \cdots,& 0)& \\
\beta_2     =& (0,&  1,& -1,& 0,& \cdots,& 0)& \\
             & \vdots  &&&&&&\\
\beta_{k-1} =& (0,&  0,& \cdots,& 0,& 1,& -1)& \\
\beta_k     =& (0,&  0,& \cdots,& 0,& 0,&  1)&
\end{array}\,\,,
\end{equation}
which is nothing but the root system of $B_k$ given in \cref{eqn:rtSO2k1}. The corresponding Lie algebra $\frak{so}(2k+1)$ is \emph{not} a subalgebra of $\frak{su}(2k)$. Nevertheless, this is the root system that generates the $C$-invariant weights. Therefore, on the odd branch $\widetilde{SU}_-(2k)$, the characters $\chi_-=\tr\left(g_-\right)$ are given by $SO(2k+1)$ characters; and the Haar measure is given by the $SO(2k+1)$ Haar measure.

The concrete character dictionary is
\begin{equation}
\chi_{-,\,\mu}^{\widetilde{SU}(2k)} = \eta_C^{} \chi_\nu^{SO(2k+1)} \,.
\label{eqn:chimG2k}
\end{equation}
Here $\mu$ denotes the highest weight of a general self-conjugate representation of $SU(2k)$:
\begin{equation}
\mu = \sum_{i=1}^{2k-1} n_i \mu_i \,,
\end{equation}
with $n_i=n_{2k-i}$, and $\mu_i$ the $SU(2k)$ fundamental weights listed in \cref{eqn:fwSUN}. The corresponding $SO(2k+1$) rep in \cref{eqn:chimG2k} is the one with the highest weight
\begin{equation}
\nu = \sum_{i=1}^{k} n_i \nu_i \,,
\end{equation}
where $\nu_i$ are the fundamental weights of $SO(2k+1)$ listed in \cref{eqn:fwSO2k1}. Note that there is also an overall intrinsic sign freedom $\eta_C^{}=\pm$ in \cref{eqn:chimG2k}, as explained before.

Taking the adjoint rep of $SU(2k)$ as an example, we have $n_1=n_{2k-1}=1$ and $n_i=0, \text{ for } i=2,\cdots,2k-2$. This tells us the corresponding $SO(2k+1)$ rep is $\nu=\nu_1$---the vector representation. Therefore, we obtain
\begin{equation}
\chi_{-,\,\text{adjoint}}^{\widetilde{SU}(2k)} = \eta_C^{} \chi_\text{vector}^{SO(2k+1)} \,.
\label{eqn:chimadjG2k}
\end{equation}
We can also verify this from the explicit character expressions. The character of the $SU(2k)$ adjoint rep is
\begin{equation}
\chi_\text{adjoint}^{SU(2k)} (y) = 2k - 1 + \sum_{i\ne j}^{2k} \frac{y_i}{y_j} \,.
\end{equation}
Here $y=\left(y_1, \cdots, y_{2k}\right)$, with $\prod_{i=1}^{2k} y_i =1$ understood (see \cref{tbl:xVariables}). Upon folding (either by average or by sum), we need to identify $y_{2k+1-i} = y_i^{-1} \text{ for } i=1, \cdots, k$ (as dictated by \cref{eqn:tildebeta}), and the above character becomes
\begin{align}
\chi_\text{adjoint-folded}^{SU(2k)} = 2k - 1 + \sum_{i=1}^k \left( y_i^2 + \frac{1}{y_i^2} \right) + 2\sum_{i<j}^k \left( \frac{y_i}{y_j} + \frac{y_j}{y_i} + y_i y_j + \frac{1}{y_i y_j} \right) \,.
\end{align}
We know that the $C$-invariant subgroup of $SU(2k)$ is $Sp(2k)$~\cite{Bourget:2018ond} (which can be figured out using folding by \emph{average}), so this folded character must be able to decompose into $Sp(2k)$ characters. Indeed, we find
\begin{equation}
\chi_\text{adjoint-folded}^{SU(2k)} = \chi_{\rho=2\rho_1}^{Sp(2k)} + \chi_{\rho=\rho_2}^{Sp(2k)} \,,
\end{equation}
with
\begin{subequations}
\begin{align}
\chi_{\rho=2\rho_1}^{Sp(2k)} &= k + \sum_{i=1}^k \left( y_i^2 + \frac{1}{y_i^2} \right) + \sum_{i<j}^k \left( \frac{y_i}{y_j} + \frac{y_j}{y_i} + y_i y_j + \frac{1}{y_i y_j} \right) \,, \\[3pt]
\chi_{\rho=\rho_2}^{Sp(2k)} &= k - 1 + \sum_{i<j}^k \left( \frac{y_i}{y_j} + \frac{y_j}{y_i} + y_i y_j + \frac{1}{y_i y_j} \right) \,,
\end{align}
\end{subequations}
Furthermore, in these two $Sp(2k)$ irreps, the charge conjugation element $C$ should just be $\pm 1$, with opposite signs. Therefore, the $C$-invariant character is given by the difference between them, with an arbitrary overall sign $\eta_C^{}=\pm 1$:
\begin{equation}
\chi_{-,\,\text{adjoint}}^{\widetilde{SU}(2k)}
= \eta_C^{} \left[ \chi_{\rho=2\rho_1}^{Sp(2k)} - \chi_{\rho=\rho_2}^{Sp(2k)} \right]
= \eta_C^{} \left[ 1 + \sum_{i=1}^k \left( y_i^2 + \frac{1}{y_i^2} \right) \right]
= \eta_{C}^{} \chi_{\nu=\nu_{1}}^{SO(2k+1)} (y^{2}) \,.
\end{equation}
This agrees with \cref{eqn:chimadjG2k} upon the redefinition $y_i \to \sqrt{y_i}$, and hence the square root in \cref{tbl:xVariables}.

\subsection{Folding $A_{2k}$}

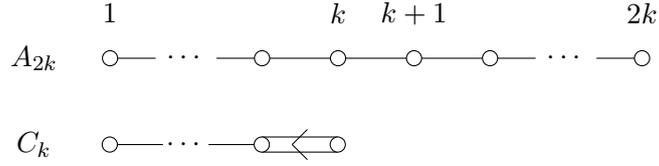
\begin{figure}
\centering
\subfigure{
\begin{tikzpicture}
	\draw (0,0) -- (0.6,0);
	\draw (1.4, 0) -- (5.6,0);
	\draw (6.4, 0) -- (7,0);	
	\draw[fill=white] (0,0) circle(.1);
	\draw[fill=white] (2,0) circle(.1);
	\draw[fill=white] (3,0) circle(.1);
	\draw[fill=white] (4,0) circle(.1);
	\draw[fill=white] (5,0) circle(.1);	
	\draw[fill=white] (7,0) circle(.1);		
	\node at (1,0) {$\cdots$};
	\node at (6,0) {$\cdots$};
	\node at (0,0.6) {$1$};		
	\node at (3,0.6) {$k$};	
	\node at (4,0.6) {$k+1$};	
	\node at (7,0.6) {$2k$};	
	\node at (-1,0) {$A_{2k}$};
\end{tikzpicture}
}\\[10pt]
\subfigure{
\begin{tikzpicture}
	\draw (0,0) -- (0.7,0);
	\draw (1.3, 0) -- (2,0);
	\draw (2,0.1) -- (3,0.1);
	\draw (2,-0.1) -- (3,-0.1);
	\draw (2.4,0) -- (2.6,0.2);
	\draw (2.4,0) -- (2.6,-0.2);		
	\draw[fill=white] (0,0) circle(.1);
	\draw[fill=white] (2,0) circle(.1);
	\draw[fill=white] (3,0) circle(.1);	
	\draw[fill=white] (7.25,0) circle(.0);	
	\node at (1,0) {$\cdots$};
	\node at (-1,0) {$C_k$};
\end{tikzpicture}
}
\caption{Folding the Dynkin diagram $A_{2k}$ by sum yields the Dynkin diagram $C_k$.}
\label{fig:A2kfolding}
\end{figure}

Let us now turn to the case $r=2k$, an even number. This one is an oddity. We normally see statements that $A_{2k}$ Dynkin diagram cannot be folded. For instance, the \href{https://en.wikipedia.org/wiki/Dynkin_diagram}{Wikipedia page on Dynkin diagram} states ``\emph{The one condition on the automorphism for folding to be possible is that distinct nodes of the graph in the same orbit (under the automorphism) must not be connected by an edge; at the level of root systems, roots in the same orbit must be orthogonal.}''  As there is no middle node in the Dynkin diagram $A_{2k}$, all the simple roots pair up under the outer automorphism, as depicted by \cref{fig:A2kfolding}. In particular, the two connected simple roots $\alpha_k$ and $\alpha_{k+1}$ have to be in the same orbit, violating the above stated condition. The folding is defined by adding columns of the Cartan matrix transformed by the automorphism. In terms of roots, it corresponds to $\alpha_{i} \rightarrow \alpha_{i} + \alpha_{2k+1-i}$. For example in the case of $A_{6}$, the folding would have produced
\begin{align}
	\left( \begin{array}{rrrrrr}
		2 & -1 & 0 & 0 & 0 & 0\\
		-1 & 2 & -1 & 0 & 0 & 0\\
		0 & -1 & 2 & -1 & 0 & 0\\
		0 & 0 & -1 & 2 & -1 & 0\\
		0 & 0 & 0 & -1 & 2 & -1\\
		0 & 0 & 0 & 0 & -1 & 2
		\end{array} \right)
	\rightarrow
	\left( \begin{array}{rrr}
		2 & -1 & 0 \\
		-1 & 2 & -1 \\
		0 & -1 & 1 \\
		0 & -1 & 1 \\
		-1 & 2 & -1 \\
		2 & -1 & 0
		\end{array} \right) \,.
\end{align}
The last three rows are clearly redundant.  Removing them, we obtain the matrix
\begin{align}
		\left( \begin{array}{rrr}
		2 & -1 & 0 \\
		-1 & 2 & -1 \\
		0 & -1 & 1
		\end{array} \right) \,,
\end{align}
which is not a legitimate Cartan matrix because $A_{3\, 3}=1 \neq 2$. This problem has been overcome in Refs. \cite{Fuchs:1996vp,Fuchs:1996ju} by allowing for an additional factor of two {for the last column, and it becomes a legitimate Cartan matrix
\begin{align}
		\left( \begin{array}{rrr}
		2 & -1 & 0 \\
		-1 & 2 & -2 \\
		0 & -1 & 2
		\end{array} \right) \,,
\end{align}
which is that of $C_{3}$.  This procedure}
generalizes to all Kac--Moody algebras. With this new definition, folding $A_{2k}$ by sum yields the following new simple roots:
\begin{equation}\renewcommand\arraystretch{1.5}
\begin{array}{rlrlrrrrrrrrrrr}
\tilde\gamma_1     =& \alpha_1 + \alpha_{2k}                &=& (1,& -1,&  0,& 0,& 0,& \cdots,& 0,& 0,& 0,&  1,& -1)& \\
\tilde\gamma_2     =& \alpha_2 + \alpha_{2k-1}              &=& (0,&  1,& -1,& 0,& 0,& \cdots,& 0,& 0,& 1,& -1,&  0)& \\
                    &\vdots                                 & & \vdots &&&&&&&&&&&\\
\tilde\gamma_{k-1} =& \alpha_{k-1} + \alpha_{k+2}           &=& (0,& \cdots,& 0,& 1,& -1,& 0,&  1,& -1,& 0,& \cdots,& 0)& \\
\tilde\gamma_k     =& 2\left(\alpha_k + \alpha_{k+1}\right) &=& (0,& \cdots,& 0,& 0,&  2,& 0,& -2,&  0,& 0,& \cdots,& 0)&
\end{array}\,\,.
\label{eqn:tildegamma}
\end{equation}
Note the additional factor of two in the last line. In terms of a root system, these are equivalent to the following set
\begin{equation}\renewcommand\arraystretch{1.3}
\begin{array}{rlrrrrrr}
\gamma_1     =& (1,& -1,&  0,& 0,& \cdots,& 0)& \\
\gamma_2     =& (0,&  1,& -1,& 0,& \cdots,& 0)& \\
             & \vdots  &&&&&&\\
\gamma_{k-1} =& (0,&  0,& \cdots,& 0,& 1,& -1)& \\
\gamma_k     =& (0,&  0,& \cdots,& 0,& 0,&  2)&
\end{array}\,\,.
\end{equation}
which is nothing but the root system of $C_k$ given in \cref{eqn:rtSp2k}. Therefore, on the odd branch $\widetilde{SU}_-(2k+1)$, the characters $\chi_-=\tr\left(g_-\right)$ are given by $Sp(2k)$ characters; and the Haar measure is given by the $Sp(2k)$ Haar measure.

The concrete character dictionary is
\begin{equation}
\chi_{-,\,\mu}^{\widetilde{SU}(2k+1)} = \eta_C^{} \chi_\rho^{Sp(2k)} \,.
\label{eqn:chimG2k1}
\end{equation}
Here $\mu$ denotes the highest weight of a general self-conjugate representation of $SU(2k+1)$:
\begin{equation}
\mu = \sum_{i=1}^{2k} n_i \mu_i \,,
\end{equation}
with $n_i=n_{2k+1-i}$, and $\mu_i$ the $SU(2k+1)$ fundamental weights listed in \cref{eqn:fwSUN}. The corresponding $Sp(2k$) rep in \cref{eqn:chimG2k1} is then the one with the highest weight
\begin{equation}
\rho = \sum_{i=1}^{k} n_i \rho_i \,.
\end{equation}
where $\rho_i$ are the fundamental weights of $Sp(2k)$ listed in \cref{eqn:fwSp2k}. Note that there is also an overall sign freedom in \cref{eqn:chimG2k1}, due to the intrinsic sign choice $\eta_C^{}=\pm$, as explained before.

Taking the adjoint rep of $SU(2k+1)$ as an example, we have $n_1=n_{2k}=1$ and $n_i=0, \text{ for } i=2,\cdots,2k-1$. This tells us the corresponding $Sp(2k)$ rep is $\rho=\rho_1$---the (first) fundamental representation. Therefore, we obtain
\begin{equation}
\chi_{-,\,\text{adjoint}}^{\widetilde{SU}(2k+1)} = \eta_C^{} \chi_\text{fundamental}^{Sp(2k)} \,.
\label{eqn:chimadjG2k1}
\end{equation}
We can also verify this from the explicit character expressions. The character of the $SU(2k+1)$ adjoint rep is
\begin{equation}
\chi_\text{adjoint}^{SU(2k+1)} = 2k + \sum_{i\ne j}^{2k+1} \frac{y_i}{y_j} \,.
\end{equation}
Here $y=\left(y_1, \cdots, y_{2k+1}\right)$, with $\prod_{i=1}^{2k+1} y_i =1$ understood (see \cref{tbl:xVariables}). Upon folding (either by average or by sum), we need to identify $y_{2k+2-i} = y_i^{-1} \text{ for } i=1, \cdots, k$ and $y_{k+1}=1$ (as dictated by \cref{eqn:tildegamma}), and the above character becomes
\begin{align}
\chi_\text{adjoint-folded}^{SU(2k+1)} = 2k + \sum_{i=1}^k \left( y_i^2 + \frac{1}{y_i^2} + 2y_i + \frac{2}{y_i} \right) + 2\sum_{i<j}^k \left( \frac{y_i}{y_j} + \frac{y_j}{y_i} + y_i y_j + \frac{1}{y_i y_j} \right) \,.
\end{align}
We know that the $C$-invariant subgroup of $SU(2k+1)$ is $SO(2k+1)$~\cite{Bourget:2018ond}, which can be figured out using folding by \emph{average}. So this folded character must be able to decompose into $SO(2k+1)$ characters. Indeed, we find
\begin{equation}
\chi_\text{adjoint-folded}^{SU(2k+1)}
= \chi_{\nu=2\nu_{1}}^{SO(2k+1)} + \chi_{\nu=\nu_{2}}^{SO(2k+1)} \,,
\end{equation}
with
\begin{subequations}
\begin{align}
\chi_{\nu=2\nu_{1}}^{SO(2k+1)} &= k + \sum_{i=1}^k \left( y_i^2 + \frac{1}{y_i^2} + y_i + \frac{1}{y_i} \right) + \sum_{i<j}^k \left( \frac{y_i}{y_j} + \frac{y_j}{y_i} + y_i y_j + \frac{1}{y_i y_j} \right) \,, \\[3pt]
\chi_{\nu=\nu_{2}}^{SO(2k+1)} &= k + \sum_{i=1}^k \left( y_i + \frac{1}{y_i} \right) + \sum_{i<j}^k \left( \frac{y_i}{y_j} + \frac{y_j}{y_i} + y_i y_j + \frac{1}{y_i y_j} \right) \,,
\end{align}
\end{subequations}
Furthermore, in these two $SO(2k+1)$ irreps, the charge conjugation element $C$ should just be $\pm 1$, with opposite signs. Therefore, the $C$-invariant character is given by the difference between them, with an arbitrary overall sign $\eta_C^{}=\pm 1$:
\begin{equation}
\chi_{-,\,\text{adjoint}}^{\widetilde{SU}(2k+1)}
= \eta_C^{} \left[ \chi_{\nu=2\nu_{1}}^{SO(2k+1)} -
	\chi_{\nu=\nu_{2}}^{SO(2k+1)} \right]
= \eta_C^{} \left[ \sum_{i=1}^k \left( y_i^2 + \frac{1}{y_i^2} \right) \right]
= \eta_{C}^{} \chi_{\rho=\rho_{1}}^{Sp(2k)} (y^{2}) \,.
\end{equation}
This agrees with \cref{eqn:chimadjG2k1} upon the redefinition $y_i \to \sqrt{y_i}$, and hence the square root in \cref{tbl:xVariables}.

\section{Hilbert series for the $p^4$ chiral Lagrangian}
\label{appsec:p4}

While our methods allow us to enumerate the chiral Lagrangian to high order, it is important to cross-check the results with known results at lower order. We discussed this for the $p^6$ Lagrangian in the main text; however, perhaps the most familiar result are the NLO terms, \textit{i.e.} the $p^4$ operators~\cite{Gasser:1983yg}. Here we provide the Hilbert series results for the $p^4$ Lagrangian, which we hope will enable readers who are familiar with the chiral Lagrangian, but less familiar with Hilbert series techniques, to gain some footing with results they already know. Throughout this appendix we will work explicitly with two flavors, \(N_f = 2\), for which there are 10 operators in the basis~\cite{Gasser:1983yg}.

One minor complication with the $p^4$ Hilbert series, which we mentioned in Sec.~\ref{subsec:IBP}, is the need for \(\D H\); that is, there are co-closed but not co-exact forms that contribute spurious information to the output of $H_0$ in Eq.~\eqref{eqn:H0} (see sec.~7 of~\cite{Henning:2017fpj} for a general, detailed discussion). In fact, the output of the Hilbert series makes it entirely obvious that something is amiss:
\begin{align}
&\left. H^{C^+P^+}_{N_f=2}\right|_{p^4} = 2u^4 + 2f_-u^2 + 2f_+ u^2 + \avg{\Sigma_+}u^2 + \avg{\Sigma_-}u^2 + 2f_+^2 + 2f_-^2 + 2f_+f_- +\Sigma_+^2 + \Sigma_-^2 \notag\\[3pt]
&\hspace{60pt} + \Sigma_+\Sigma_- + \avg{\Sigma_+}^2 + \avg{\Sigma_-}^2 + \avg{\Sigma_+}\avg{\Sigma_-} + \textcolor{blue}{D^4 - Df_+u - Df_-u - Du^3} \,,
\end{align}
where we have highlighted in blue the terms which are seemingly non-sensical (there is no operator composed of only four derivatives, and the other terms have negative coefficients). In fact, it's easy to explicitly identify the co-closed but not co-exact forms that lead to the issue:
\begin{equation}\label{eqn:DeltaHForms}
\def\arraystretch{1.6}
\setlength{\arrayrulewidth}{.3mm}
  \begin{array}{c|c} \text{form} & \hspace{5mm}\text{contribution to }H_0\hspace{5mm} \\
    \hline
    \e^{\m\n\r\s} & +D^4 \\
    \e_{\m\n\r\s}f_+^{\n\r}u^{\s} & -Df_+u \\
    \e_{\m\n\r\s}f_-^{\n\r}u^{\s} & -Df_-u \\
    \hspace{5mm}\e_{\m\n\r\s}u^{\n}u^{\r}u^{\s}\hspace{5mm} & -Du^3
  \end{array}
\end{equation}
As explained in~\cite{Henning:2015alf,Henning:2017fpj}, the full Hilbert series can be written as \(H = H_0 + \Delta H\), where \(\Delta H\) contains the information about co-closed but not co-exact forms. Similar to the different branches of \(H_0\) defined in \cref{eqn:HBranches}, we find different branches for \(\D H\) determined from the various quantum numbers of the operators above in \cref{eqn:DeltaHForms}:
\begin{subequations}
\begin{align}
\left. \D H^{C^+P^+}_{N_f = 2} \right|_{p^4} &= - D^4 + D f_+u + Df_-u + Du^3 \,, \\[8pt]
\left. \D H^{C^+P^-}_{N_f = 2} \right|_{p^4} &= + D^4 + D f_+u - Df_-u + Du^3 \,, \\[8pt]
\left. \D H^{C^-P^+}_{N_f = 2} \right|_{p^4} &= - D^4 - D f_+u + Df_-u - Du^3 \,, \\[8pt]
\left. \D H^{C^-P^-}_{N_f = 2} \right|_{p^4} &= + D^4 - D f_+u - Df_-u - Du^3 \,.
\end{align}
\end{subequations}
These can be combined analogously to Eqs.~\eqref{eqn:HcasesBasic} and~\eqref{eqn:HcasesMore}; here we list only a few such combinations:
\begin{subequations}
\begin{align}
\left. \D H^{C\text{-even}\, P\text{-even}}_{N_f=2} \right|_{p^4} &= 0 \,, \\[8pt]
\left. \D H^{CP\text{-even}}_{N_f=2} \right|_{p^4} &= 0 \,, \\[8pt]
\left. \D H^{CP\text{-odd}}_{N_f=2} \right|_{p^4} &= -D^4 + Df_+u + Df_-u + Du^3 \,.
\end{align}
\end{subequations}

Accounting for the above, we arrive at the \(C\)-even, \(P\)-even Hilbert series for the \(p^4\) chiral Lagrangian,
\begin{equation}
\left. H^{C\text{-even}\, P\text{-even}}_{N_f=2} \right|_{p^4} = 2u^4 + f_+u^2 + \avg{\Sigma_+}u^2 + f_+^2 + f_-^2 + \Sigma_+^2 + \Sigma_-^2 + \avg{\Sigma_+}^2 + \avg{\Sigma_-}^2 \,,
\end{equation}
which, indeed, tells us that there are ten operators in the $p^4$ basis~\cite{Gasser:1983yg}. For fun, we also list out the \(CP\)-even and \(CP\)-odd results:
\begin{subequations}
\begin{align}
\left. H^{CP\text{-even}}_{N_f=2} \right|_{p^4} &= 2u^4 + f_+u^2 + f_- u^2 + \avg{\Sigma_+}u^2 + f_+^2 + f_-^2 + f_+f_-+ \Sigma_+^2 + \Sigma_-^2 \notag\\
&\quad + \avg{\Sigma_+}^2 + \avg{\Sigma_-}^2 \,, \\[8pt]
\left. H^{CP\text{-odd}}_{N_f=2} \right|_{p^4} &= f_+u^2 + f_- u^2 + \avg{\Sigma_-}u^2 + f_+^2 + f_-^2 + f_+f_-+ \Sigma_+\Sigma_- + \avg{\Sigma_+}\avg{\Sigma_-} \,.
\end{align}
\end{subequations}


\section{Hilbert series for anomalous terms in the $p^8$ chiral Lagrangian}
\label{appsec:p8}

In the attached auxiliary material we include a {\tt Mathematica} notebook containing the full Hilbert series for the anomalous $C$-even \emph{and} $P$-even chiral Lagrangian at chiral dimension $p^8$. In this appendix, we provide a breakdown of the enumeration of the classes of operators appearing at this order, which mirrors the breakdown of the non-anomalous terms that appeared in Tables 3-8 of Ref.~\cite{Bijnens:2018lez}. In particular, we consider four cases:
\begin{enumerate}
\item All fields included
\item Excluding scalar and pseudo-scalar fields $\Sigma_\pm$, $\langle\Sigma_\pm\rangle$
\item Excluding vector and axial-vector fields $f_{\pm\mu\nu}$
\item Excluding all the external fields $\Sigma_\pm$, $\langle\Sigma_\pm\rangle$, and $f_{\pm\mu\nu}$
\end{enumerate}
In \cref{tbl:breakdown}, we list the total number of operators in each of these cases, for $N_f=2$, $N_f=3$, and the general $N_f$ case (operationally $N_f\ge8$ in our approach). For more detailed breakdowns, we refer the reader to the auxiliary material.

\begin{table}[t]
\renewcommand{\arraystretch}{1.6}
\setlength{\arrayrulewidth}{.3mm}
\setlength{\tabcolsep}{1em}
\begin{center}
\begin{tabular}{c|ccc}
 Anomalous $p^8$                               & $N_f$ & $N_f=3$ & $N_f=2$ \\\hline
 All fields                                    &   999 &     705 &      92  \\
 No $\Sigma_\pm$ or $\langle\Sigma_\pm\rangle$ &   565 &     369 &       0  \\
 No $f_{\pm\mu\nu}$                            &    79 &      45 &       2  \\
 Only $u_\mu$                                  &    36 &      16 &       0  \\
\end{tabular}\vspace{0.5cm}
\caption{Breakdown of the anomalous operators at $p^8$ in the chiral Lagrangian.}
\label{tbl:breakdown}
\end{center}
\end{table}

\section{Enumeration of operators for $N_f=4,5$ up to chiral dimension 16}
\label{appsec:nf45}

The following table contains the enumeration of operators used for the $N_f=4$ and $N_f=5$ curves shown in Fig.~\ref{fig:growth}.

\begin{center}
{\small
\begin{tabular}{ccccccccc}
 Chiral Dim: & $p^2$ & $p^4$ & $p^6$ & $p^8$& $p^{10}$ & $p^{12}$ &$p^{14}$ & $p^{16}$\\
\hline
SU(4) & 2 & 13 & 136 & 2702 & 78632 & 2675469 & 95181455 & 3419764470 \\
SU(5) &  2 & 13 & 138 & 2837 & 88575 & 3346187 & 135986333 & 5710835325 \\
\end{tabular}
}
\end{center}

\bibliographystyle{JHEP}
\bibliography{chpt}

\end{document}